\documentclass[fleqn,usenatbib]{mnras}

\usepackage{newtxtext,newtxmath}

\usepackage[T1]{fontenc}
\usepackage{xcolor}

\DeclareRobustCommand{\VAN}[3]{#2}
\let\VANthebibliography\thebibliography
\def\thebibliography{\DeclareRobustCommand{\VAN}[3]{##3}\VANthebibliography}

\usepackage{graphicx}	

\title[Constraining ULDM with the NSC]{Constraining Ultra Light Dark Matter with the Galactic Nuclear Star Cluster}

\author[F. Toguz et al.]{
Firat Toguz$^{1}$\thanks{E-mail: firat.toguz.19@ucl.ac.uk},
Daisuke Kawata$^{1}$,
George Seabroke$^{1}$,
Justin I. Read$^{2}$
\\
$^{1}$Mullard Space Science Laboratory, University College London, Holmbury St. Mary, Dorking, Surrey, RH5 6NT, UK \\
$^{2}$ Department of Physics, University of Surrey, Guildford, GU2 7XH, UK 
}

\date{Accepted XXX. Received YYY; in original form ZZZ}

\pubyear{2015}

\begin{document}
\label{firstpage}
\pagerange{\pageref{firstpage}--\pageref{lastpage}}
\maketitle

\begin{abstract}
We use the Milky Way's nuclear star cluster (NSC) to test the existence of a dark matter `soliton core', as predicted in ultra-light dark matter (ULDM) models. Since the soliton core size is proportional to $m_{\rm DM}^{-1}$, while the core density grows as $m_{\rm DM}^{2}$, the NSC (dominant stellar component within $\sim 3$\,pc) is sensitive to a specific window in the dark matter particle mass, $m_{\rm DM}$. We apply a spherical isotropic Jeans model to fit the NSC line-of-sight velocity dispersion data, assuming priors on the precisely measured Milky Way's supermassive black hole (SMBH) mass and the well-measured NSC density profile. We find that the current observational data reject the existence of a soliton core for a single ULDM particle with mass in the range $10^{-20.4}$~eV $\lesssim m_{\rm DM}$ $\lesssim 10^{-18.5}$~eV, assuming that the soliton core structure is not affected by the Milky Way's SMBH. We test our methodology on mock data, confirming that we are sensitive to the same range in ULDM mass as for the real data. Dynamical modelling of a larger region of the Galactic centre, including the nuclear stellar disc, promises tighter constraints over a broader range of $m_{\rm DM}$. We will consider this in future work.
\end{abstract}




\section{Introduction}
\label{sec:intro}

The $\Lambda$ cold dark matter ($\Lambda$CDM) cosmological model successfully describes the cosmic microwave background (CMB) \citep[e.g.][]{Bennet2013, Planck2020} and large scale structure \citep[e.g.][]{Percival2001, Tegmark2004, Weinberg2015}. However, tensions between theory and observations persist at small scales \citep[e.g.][for a review]{BullockBoylan2017}. One example is the ``missing satellite" problem \citep[e.g.][]{klypin1999, moore1999}, in which numerical simulations of a Milky Way-like galaxy in $\Lambda$CDM predict that $\sim$ a thousand dark matter subhalos large enough to host a visible galaxy ($M_{\rm halo}>\sim10^{7} M_{\odot}$) should be found orbiting within the Milky Way. 
However, to date only $\sim 70$ satellite dwarf galaxies have been found \citep[e.g.][]{Drlica-Wagner+Bechtol+Mau+20}.
Another example is the `cusp-core problem' \citep[e.g.][]{flores1994, moore1994}. Pure dark matter $N$-body simulations of structure formation in $\Lambda$CDM predict that bound dark matter halos have a centrally divergent `cuspy' density profile \citep{navarro1997}. By contrast, observations of the the rotation curves of nearby low-surface brightness galaxies favour instead a much lower density inner `core' \citep[e.g][]{de2001}.

The above small scale puzzles may owe entirely to `baryonic effects' (i.e. due to gas cooling, star formation and stellar feedback) not included in early structure formation models. Galaxy formation is expected to become increasingly inefficient at low mass due to a combination of stellar feedback and ionising radiation from the first stars \citep[e.g.][]{1992Efstathiou, 2002Benson,2016Sawala}. Indeed, recent dynamical estimates of the masses of the Milky Way's dwarf companions suggests that there is no missing satellite problem at least down to a halo mass of $M_{200} \sim 10^9$\,M$_\odot$  \citep{2019ErkalRead}. Furthermore, repeated gas inflow/outflow, driven by gas cooling and stellar feedback, can cause the central gravitational potential in dwarf galaxies to fluctuate with time. This pumps energy into the dark matter particle orbits causing the halo to expand \citep{1996Navarro,2005ReadGilmore,2012PontzenGovernato,DiCintio+Brook+Maccio+14}. There is mounting observational evidence that this `dark matter heating' effect has occurred in nearby dwarf galaxies; this may be sufficient to fully solve the cusp-core problem \citep[e.g.][]{2019Read}.

Nonetheless, $\Lambda$CDM's small scale puzzles have inspired a host of novel dark matter models designed to lower the inner density of dark matter halos and/or reduce the number of dark matter subhalos. These include warm dark matter \citep[WDM e.g.][]{1994Dodelson, 2001Bode} and ultra-light dark matter \citep[ULDM e.g.][]{ferreira2020ultra, 2021Hu}. In WDM, dark matter is assumed to be relativistic for a time in the early Universe, suppressing the small scale power spectrum and leading to fewer, lower-density, satellite galaxies as compared to CDM. This can naturally occur if, for example, dark matter is a light thermal relic particle. 

For thermal relic masses of about $\sim 1\,$keV, WDM has the potential to resolve the missing satellite problem \citep[e.g.][]{2002Knebe, 2021Lovell, 2014Lovell}, although this depends on the assumed total mass of the Milky Way \citep[e.g.][]{2014Kennedy}. Indeed, the observed number of the Milky Way satellite galaxies puts a lower limit of the WDM mass \citep[e.g.][]{2011Polisensky}. \citet{2020Newton} favour a lower limit of 3.99\,keV, marginalising the uncertainty in the Milky Way mass, and taking into account the expected inefficiency of dwarf galaxy formation \citep[see also an even stronger constraint of $>6.5$~keV in][]{2021Nadler}. A  similar lower limit on the WDM mass is imposed by the other astronomical probes, such as Lyman-$\alpha$ forest data \citep[]{2017Ir}, strong gravitational lensing \citep[]{2020Gilman} and density fluctuations in Galactic stellar streams \citep[]{2019Banik}. However, $\sim$ keV-scale WDM is not able to solve the cusp-core problem on its own \citep[see e.g.][for a review]{Weinberg2015}. \citet{2012Macci} show that a WDM mass of about 0.1~keV is required to generate $\sim $kpc-sized cores in dwarf galaxies, but such a low mass WDM particle is incompatible with the above observational constraints. 
    
ULDM has emerged as a novel dark matter model that can solve both the cusp-core and missing satellite problems on its own, without recourse to baryonic effects. ULDM is a type of dark matter that is made up of bosons with mass in the range $10^{-22.0}~{\rm eV} < m_{\rm DM} <1~{\rm eV}$ \citep[e.g.][for a review]{ferreira2020ultra,2021Hu}.
On large scales, ULDM behaves just like CDM, i.e. it successfully explains large scale structure and the CMB. However, in high density regions like the centres of dark matter halos, the de Broglie wavelength of the ULDM particles becomes larger than the mean inter-particle separation, and the ULDM undergoes Bose-Einstein condensation. Consequently, ULDM introduces a new scale length -- the Jeans length, $\lambda_J$ -- set by the de Broglie wavelength and the dark matter density \citep{2000hu}:

\begin{eqnarray}
    \lambda_{\rm J}
    &\sim& 55[m_{\rm DM}/(10^{-22}~{\rm eV})]^{-1/2}(\rho/\rho_{\rm b})^{-1/4}\nonumber \\ &\times&(\Omega_{\rm ULDM}h^{2})^{-1/4}~{\rm kpc},
    \label{eq:Jeansscale}
\end{eqnarray}
where $\rho$ is the matter density, $\Omega_{\rm ULDM}$ is the mass fraction for the ULDM particle with respect to the critical density, and $\rho_{\rm b}\sim 2.8\times10^{11}(\Omega_{\rm ULDM}h^{2})$~M$_{\sun}$~Mpc$^{-3}$ is the background density.

Perturbations larger than $\lambda_J$ will collapse similarly to CDM, while perturbations smaller than $\lambda_J$ are stabilized by quantum pressure due to the uncertainty principle \citep[e.g.][]{Hu2000}. At low dark matter density, close to the background density of the Universe, the Jeans mass can be computed from the Jeans length, as follows \citep[e.g.][]{2017hui}:
\begin{eqnarray}
    M_{\rm J} &=& \frac{4\pi}{3}\rho\left(\frac{1}{2}\lambda_{\rm J}\right)^{3} \nonumber \\
    &\simeq& 1.5 \times 10^{7}\rm M_{\odot}(1 + z)^{3/4}\left(\frac{\Omega_{\rm ULDM}}{0.27}\right)^{1/4}\nonumber \\
    &\times&\left(\frac{H_0}{70~{\rm km~s^{-1}~Mpc^{-1}}}\right)^{1/2}\left(\frac{10^{-22} {\rm eV}}{m}\right)^{3/2},
\end{eqnarray}
where $H_0$ is the Hubble constant. This Jeans mass corresponds to the minimum halo mass which can collapse in the ULDM model; it leads to a smaller number of dwarf galaxies as compared to the CDM model. In this way, ULDM can resolve the missing satellite problem \citep[e.g.][]{2020Kulkarni}. According to \citet{2021Nadler}, the observed number of Milky Way satellites requires a ULDM particle mass higher than $2.9\times10^{-21.0}$~eV.

Another consequence of ULDM is that, at the scale of the de Broglie wavelength within the collapsed halo, the Bose-Einstein condensation develops a `soliton core' at the centres of galaxies \citep[e.g.][]{Hu2000,schive2014}. The soliton core has a half-mass radius of about 300\,pc in a $M_{200} \sim 10^{9}$~M$_{\odot}$ dwarf galaxy halo for a ULDM model with $m_{\rm DM}=10^{-22.0}$~eV (see eq.~(\ref{Eq:Solitondenisty}) in Sec.~\ref{subsection:DMdensity}). This soliton core can mitigate the cusp-core problem. \citet{schive2014} suggest that $m_{\rm DM}=8 \times 10^{-23.0}$~eV ULDM can explain the observed mass profile of the Fornax dwarf galaxy \citep[e.g.][]{amorisco2013,2019Read}. However, \citet{safarzadeh2020} argued that no single ULDM particle mass can explain the current observations of the ultra-faint dwarfs and the Fornax and Sculptor dwarf spheroidal galaxies simultaneously \citep[see also][]{2021Hayashi}, unless the baryonic physics changes the density profile of the dark matter halo (see above) or the observational constraints are relaxed. As summarised in Fig.~3 of \citet{2021Hayashi}, taken at face value, no single particle ULDM model can satisfy all current observational constraints, including the Lyman-alpha forest limit of $m_{\rm DM}>10^{-21.0}$~eV \citep[e.g.][]{2017Kobayashi}. Also, \citet{2019Desjacques} suggested that the black hole-halo mass relation of galaxies rules out $m_{\rm DM}<10^{-18.0}$~eV.  Thus -- at least as a full solution to $\Lambda$CDM’s small scale puzzles -- ULDM appears to be on the ropes. However, all of the current constraints on ULDM come with their own potential systematics. As such, independent observational constraints are invaluable in determining once and for all whether we can discard ULDM as a full solution to $\Lambda$CDM’s small scale puzzles.

In this paper, we consider whether the Milky Way's Nuclear Star Cluster (NSC) can provide a new and competitive probe of ULDM models. Due to it being only about 8\,kpc away from us, the stellar kinematics of the central region of the Milky Way can be more precisely measured than for more distant dwarf galaxies (d$\sim$100 kpc). Hence, the inner gravitational potential of the Milky way can be derived from precise measurements of the stellar kinematics and density distribution of tracer stars in the centre of the Galaxy. 

The Milky Way's NSC is a dense and massive star cluster \citep[NSC, e.g.][for a review]{bland2016} that harbours the Milky Way's supermassive black hole (SMBH), called ``Sgr A*'' \citep[e.g.][]{genzel1996, ghez2008}. The SMBH mass, $M_{\rm BH} = 4.261$ $\pm$ $0.012 \times 10^{6}$~M$_{\odot}$, is now precisely measured by the GRAVITY collaboration \citep{abuter2020detection}, a cryogenic, interferometric beam combiner of all four UTs of the ESO VLT with adaptive optics. The mass of the NSC itself is about $10^{7}$~M$_{\odot}$ \citep[e.g.][]{bland2016, chatzopoulos2015,Feldmeier-Krause+Zhu+Neumayer+17}. The majority ($\sim80\%$) of the stellar mass of the NSC formed more than 5\,Gyrs ago \citep[e.g.][]{gallego2018}. Thus, we can expect that the NSC is dynamically relaxed and, therefore, a good target for equilibrium mass modelling \citep[e.g.][]{Binney+Tremaine08}.

The number density of NSC stars dominate over other Milky Way stellar components up to about 3\,pc \citep[e.g.][]{gallego2018,Gallego-Cano+Schodel+Nogueras-Lara+20}. As such, we can assume that almost all of the stars observed within 3\,pc from the Milky Way's SMBH are NSC stars, and use these to trace the inner dynamical mass profile of the Galactic centre. In ULDM models, the dark matter mass profile on this small scale can be affected by the soliton core if the ULDM mass is less than about $10^{-19.0}$~eV, as suggested by Fig.~15 of \citet{bar2018}. Hence, a dynamical model of the NSC promises a new and competitive probe of ULDM. Taking advantage of the recent precise measurement of the Milky Way's SMBH mass, and the density profile of the NSC, in this paper we study if a ULDM soliton core can be detected or rejected by the existing kinematical data for NSC stars, as measured by \citet{fritz2016}. \citet{2019Bar2} excluded $2\times10^{-20.0} < m_{\rm DM}<8\times10^{-19.0}$~eV from the stellar dynamics around Sgr A* (<$\sim$0.3 pc) of the Milky Way. Our study is expected to provide a stronger constraint using the NSC stellar kinematics within about 3 pc. 

This paper is organised as follows: In Section~\ref{sec:method}, we describe the observational data and our fitting methodology. In Section~\ref{sec:results}, we describe our results. In Section~\ref{sec:Mock}, we use mock data to test the voracity of our results. Finally, in Section~\ref{sec:conclusion} we present our conclusions. Throughout this paper, we consider that dark matter consists of a single mass ULDM particle.

\section{Method}
\label{sec:method}
To derive the total mass distribution in the NSC, we use a spherically symmetric and isotropic dynamical model. Because the NSC is dominant only within about 3~pc \citep{gallego2018}, we focus on the mass distribution within 3~pc in this paper. The structure of the NSC is not a perfect sphere, it is rather a flattened sphere with a minor to major axis ratio of $q = 0.80 \pm 0.04$ \citep{fritz2016}. However, in this paper we consider that the NSC is nearly spherical, and can be approximated, therefore, by a spherical model \citep[e.g.][]{read2017break}. \citet{fritz2016} used the projected radial and tangential velocity dispersion from the proper motions of the NSC stars to show that the NSC is close to isotropic. Hence, we also assume the NSC stellar kinematics are isotropic. Using the spherical isotropic Jeans equation, we can derive the total mass of the Galactic centre as a function of the 3D radius, $r$, from the surface density profile and projected velocity dispersion profile of the stars within the NSC. Although \citet{fritz2016} also provides the proper motions of the NSC stars, we use only the line-of-sight velocity dispersion because we assume an isotropic spherical model and the uncertainties of the line-of-sight velocities are clearly defined, while the uncertainties of the tangential velocities from the proper motions are difficult to be properly assess due to their dependence on the unknown distances.
The components of the Galactic centre that affect the stellar kinematics are the SMBH, NSC and any central dark matter, including a soliton core if the correct dark matter model is ULDM. The total mass, $M_{\rm tot}(<r)$, in the Galactic centre is given by $M_{\rm tot}(<r) = M_{\rm BH} + M_{\rm NSC}(<r) + M_{\rm DM}(<r)$\footnote{There is a circumnuclear gas disc within $\sim$3~pc, whose mass could be as large as $10^6$~M$_{\sun}$ \citep[e.g.][]{2005Christopher}. Since the estimate of the gas mass is uncertain, and this mass is about 10~\% of our derived total mass within 3~pc, we do not include the contribution of the gas component to the total potential. This simplification makes more room for the ULDM soliton core to contribute the total mass, which leads to more conservative bounds on the ULDM particle mass.}. 

We adopt the recently precisely measured mass of the SMBH of $M_{\rm BH}= 4.261\pm0.012\times 10^{6}$~M$_{\odot}$ \citep{abuter2020detection} as a strong prior (see Sec.~\ref{subsection:fitting}). \citet{abuter2020detection}
note that the systematic uncertainty is larger than this statistical uncertainty. In  Appendix~\ref{sec:appendix}, we demonstrate that the results of this paper do not change if the
black hole mass is varied over this larger systematic uncertainty of about $0.06\times10^{6}$~M$_{\sun}$. 
The stellar mass of the NSC within $r$, $M_{\rm NSC}(<r)$, can be computed from the observed stellar number density profile, fitting a constant stellar mass and number density ratio. Although a CDM halo \citep{navarro1997} provides a negligible mass contribution within the NSC ($<0.1$\%), if the dark matter is ULDM, with a particle mass of around $10^{-20.0}$~eV, there should be a significant contribution of the soliton core of ULDM within the NSC. In the following subsections, we describe Jeans equation (Sec.~\ref{subsection:JeansE}), the velocity dispersion data of the NSC (Sec.~\ref{subsection:NSCvdispdata}), the stellar density profile of the NSC (Sec.~\ref{subsection:NSCdensity}), our ULDM model (Sec.~\ref{subsection:DMdensity}), and our fitting methodology (Sec.~\ref{subsection:fitting}).
\begin{figure}
    \centering
    \includegraphics[width = \columnwidth]{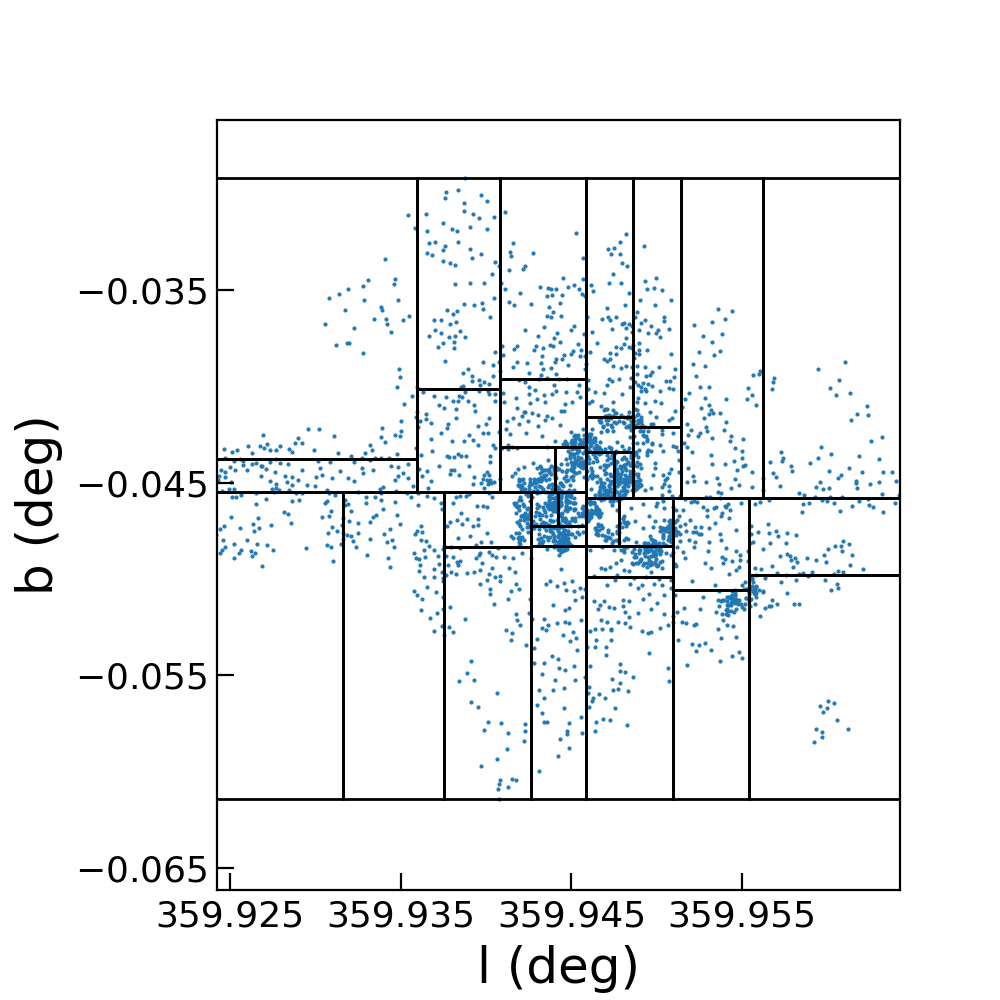}
    \caption{The distribution of stars whose line-of-sight velocities are measured in \citet{fritz2016}. The data are decomposed into 32 bins, with approximately 79 stars per bin.}
    \label{fig:KDTree}
\end{figure}
\subsection{Jeans Equation}
\label{subsection:JeansE}

For a steady state spherical stellar system that is isotropic, the Jeans equation is given by \citep[e.g.][]{Binney+Tremaine08}:
\begin{equation}
    \frac{1}{n(r)}\frac{\partial(n(r)\sigma(r)^{2})}{\partial{r}}  = - \frac{GM_{\rm tot}(<r)}{r^{2}},
    \label{eq:Jeans}
\end{equation}
where $\sigma(r)$ is the velocity dispersion of stars in NSC, $n(r)$ is the 3D number density profile of NSC stars, and $M_{\rm tot}(<r)$ is the enclosed total mass of the system within $r$. 

Integrating both sides of equation (\ref{eq:Jeans}) gives a velocity dispersion profile of:
\begin{equation}
    \sigma(r) = \sqrt{\frac{1}{n(r)}\int^{\infty}_{r}\frac{GM_{\rm tot}(<r)n(r)}{r^{2}}dr}.
    \label{eq:Jeansvdisp}
\end{equation}
Through an Abel transformation of equation (\ref{eq:Jeansvdisp}), the line-of-sight velocity is derived as:
\begin{equation}
    \sigma_{\rm LOS}(R) = \sqrt{\frac{2}{\Sigma(R)}\int^{\infty}_{R}\frac{n(r)\sigma^{2}_{r}(r)r}{\sqrt{r^{2} - R^{2}}}dr},
    \label{eq:vdisplos}
\end{equation}
where $R$ is the projected 2D radius, and $\Sigma(R)$ is the projected NSC surface number density profile, which is given by:
\begin{equation}
  \Sigma(R) = 2\int^{\infty}_{R}\frac{r n(r)}{\sqrt{r^{2} - R^{2}}}dr.
  \label{eq:surfden}
\end{equation}

\subsection{Velocity Dispersion Data}
\label{subsection:NSCvdispdata}

We use the line-of-sight velocity data measured by \citet[]{fritz2016} with the integral field spectrometer, {\it VLT/SINFONI}. \citet[]{fritz2016} obtained the line-of-sight velocities for 2,513 late-type giant stars within $R<95\arcsec$ from Sgr~A$^{*}$. Note that in this paper, we use the notation $r$ for the 3D spherical radius, and $R$ for the projected 2D radius from Sgr~A$^{*}$. The distribution of stars whose line-of-sight velocities are provided by \citet[]{fritz2016} is shown in Galactic coordinates in Fig.~\ref{fig:KDTree}.
We use a KD-Tree decomposition to bin the data (Fig.~\ref{fig:KDTree}), so that there are 32 bins, and each bin has about 79 stars. We found that this is a good compromise to maximise the number of bins, but minimise the Poisson noise in each bin. 

For the sample of stars in each bin, the line-of-sight velocity dispersion is normally computed using the following formula:
\begin{equation}
    \sigma = \sqrt{<v_{\rm LOS}^{2}> - <v_{\rm LOS}>^{2}},
\label{eq:normalvdisp}
\end{equation}
where $v_{\rm LOS}$ is the line-of-sight velocity of the star. Following \citet[]{fritz2016}, to take into account the contribution of the rotation approximately, we instead use:
\begin{equation}
    \sigma_{\rm LOS} = \sqrt{<v_{\rm LOS}^{2}>},
\end{equation}
i.e. ignoring  $<v_{\rm LOS}>^{2}$ in equation~(\ref{eq:normalvdisp}). This is based on the approximation often used as effective velocity dispersion in the kinematical analysis of the external galaxies \citep[e.g.][]{2009GultekinKayhan}, where $<v_{\rm LOS}>$ corresponds to the projected rotation curve and from equation~(\ref{eq:normalvdisp}), $<v_{\rm LOS}^{2}>$ = $\sigma^{2}$ + $<v_{\rm LOS}>^{2}=\sigma^2+v_{\rm rot}^2$, considering the kinetic energy being proportional to $\sigma^{2}$ + $<v_{\rm LOS}>^{2}$ \cite[]{Binney+Tremaine08}.

We find that the mean uncertainty of the velocity dispersion measurements from the observational errors of line-of-sight velocities is about 1.7~km~s$^{-1}$, which is smaller than the mean Poisson error of about 8~km~s$^{-1}$. For this reason, we assume that the error on each bin owes solely to the Poisson error. Following \citet[]{fritz2016}, we measure the Poisson error of the velocity dispersion with $\sigma_{\rm LOS,err,i}=\sigma_{\rm LOS,fit}(R_{\rm i})/\sqrt{2N_{\rm i}}$, where $N_{\rm i}$ is the number of stars in $i$-th bin and $R_{\rm i}$ is the mean projected radius of the stars in $i$-th bin. $\sigma_{\rm LOS,fit}(R)$ is the fitted 3rd order polynomial velocity dispersion profile. Because $\sigma_{\rm LOS,err}$ changes depending on $\sigma_{\rm LOS,fit}(R)$, we iteratively derive $\sigma_{\rm LOS,err,i}$.

We compute the line-of-sight velocity dispersion and uncertainties as described above, which are plotted against the mean radius of the stars within each bin in Fig.~\ref{fig:MODEL1SAMPLE}. We fit these observed velocity dispersion with the model described in Section~\ref{subsection:JeansE}. 

\subsection{NSC Density profiles}
\label{subsection:NSCdensity}

Following \citet{gallego2018}, we describe the 3D density profile, $\rho_{\rm NSC}(r)$, of the NSC with a 3D Nuker law \cite[]{1995Lauer}: 
\begin{equation}
    \rho_{\rm NSC}(r) = \rho_{\rm b,NSC}2^{(\beta - \gamma)/\alpha}\left(\frac{r}{r_{\rm b}}\right)^{-\gamma}\left[1 + \left(\frac{r}{r_{\rm b}}\right)^{\alpha}\right]^{(\gamma - \beta)/\alpha},
    \label{eq:nuker3D}
\end{equation}
where $r_{\rm b}$ is the break radius, $\rho_{\rm b,NSC}=\rho_{\rm NSC}(r_{\rm b})$ is the mass density of the NSC at the break radius, $\gamma$ and $\beta$ are the exponent of the inner and outer power-law slope, respectively, and $\alpha$ describes the sharpness of the transition between the inner and outer power-law profiles. \citet{gallego2018} fit the NSC stellar distribution from the high-resolution near-infrared photometric data with the 2D projected density profile of equation~(\ref{eq:nuker3D}). We rely on the precise measurement of the NSC density profile from \citet{gallego2018}, and when we fit the velocity dispersion, we fix the density profile parameters with their best fit profile.

\citet{gallego2018} demonstrate that the NSC number density profile depends on the selection of the observational data, which indicates the systematic uncertainties of the measurements of the density profile of the NSC. We take one of the best fitting models from \citet{gallego2018}: $\alpha$ = 10, $\beta$ = 3.4, $\gamma$ = 1.29 and $r_{\rm b}$ = 4.3 pc \citep[ID10 of Table 5 in][]{gallego2018}. This is the case that excludes contamination from pre-main sequence stars. We consider this to be most appropriate for our kinematic sample, since the kinematic data of \citet{fritz2016} are for late-type giants. This model also leads to the smallest $\gamma$ value, allowing for the maximal amount of dark matter within the NSC and, thereby, ensuring maximally conservative constraints on the ULDM mass. However, we tested also a value of $\gamma=1.43$, taken from a different best-fitting model from \citet{gallego2018}, and find that our results are not sensitive to these choices.
 
Although the stellar number density profile is well observed by \citet{gallego2018}, we need to convert it to the mass density profile to obtain the NSC mass contribution to the gravitational potential in the Jeans equation. Because the mass to light ratio of the observed stars are uncertain, we adopt $\rho_{\rm b,NSC}$ as a parameter when fitting the velocity dispersion profile, and marginalise over the mass scaling of the density profile. To take into account the observational uncertainty of the number density profile, we also take $\gamma$, which controls the profile in the radial range of our interest, as a fitting parameter with the prior of $\gamma=1.29\pm0.05$ \citep{gallego2018}.
\begin{figure}
    \centering
    \includegraphics[width = \columnwidth]{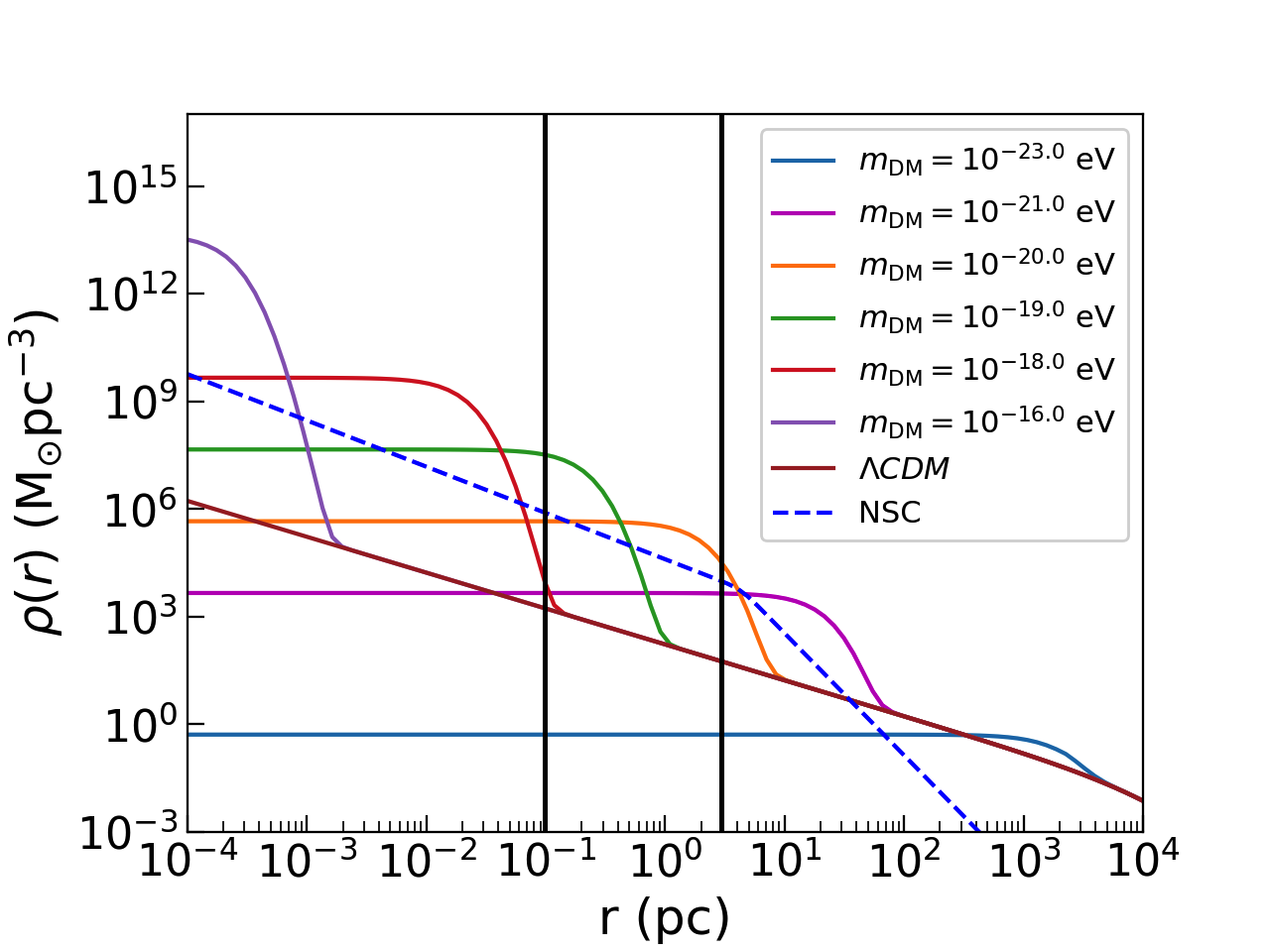}
    \caption{The density profile for the Milky Way's NSC (blue dashed), and for the central dark matter density assuming ${\Lambda}$CDM (brown) and dark matter with a ULDM particle mass of $10^{-23.0}$~eV (yellow), $10^{-21.0}$~eV (magenta), $10^{-20.0}$~eV (orange), $10^{-19.0}$~eV (green), $10^{-18.0}$~eV (black) and $10^{-16.0}$~eV (red). Notice that over the fixed radial range probed by the NSC stellar kinematic data (vertical black lines), only ULDM models with mass in a specific range will affect the stellar kinematics.
    }
    \label{fig:densityprofiles}
\end{figure}
\begin{figure}
    \centering
    \includegraphics[width = \columnwidth]{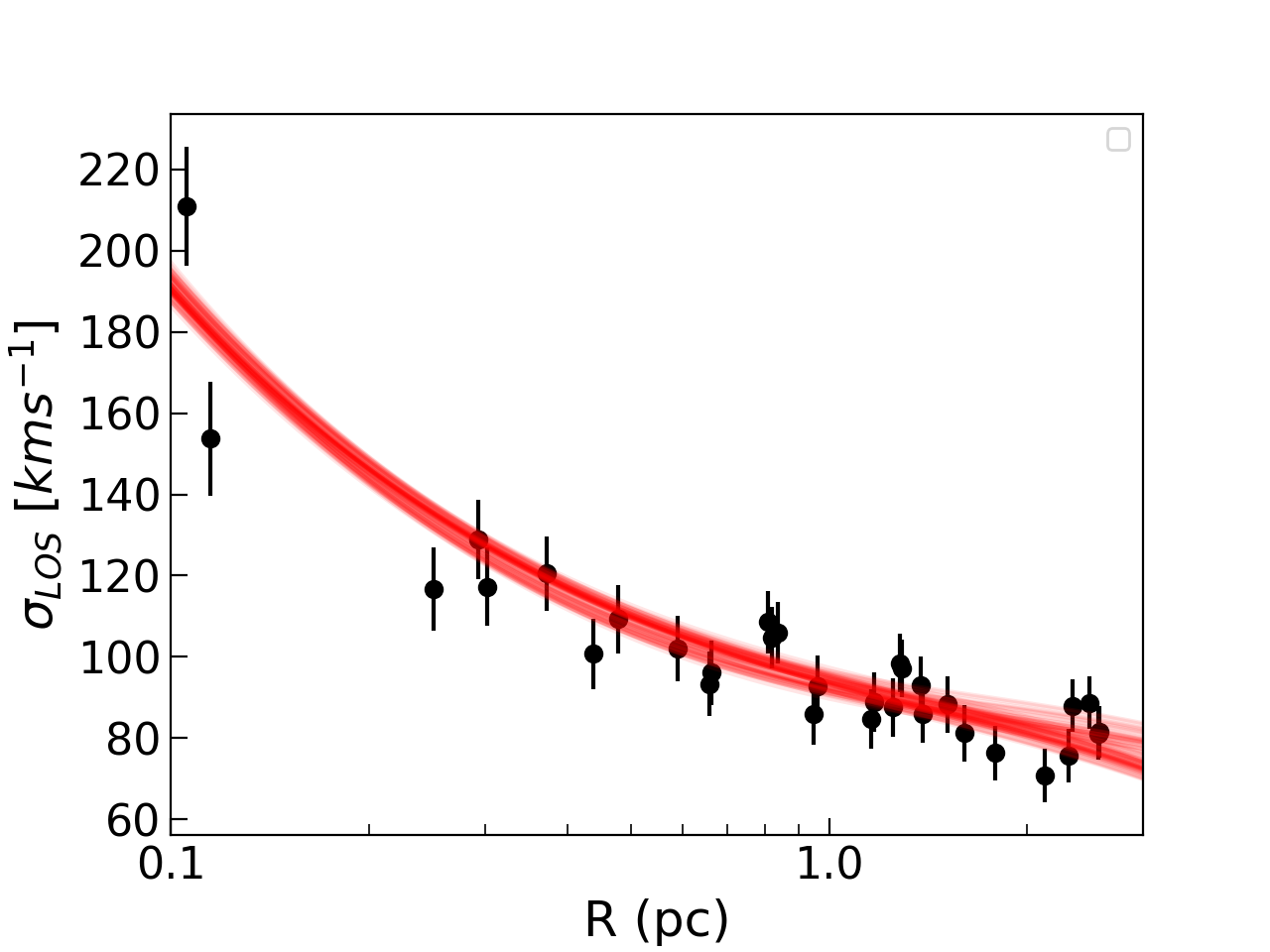}
    \caption{The observed line-of-sight velocity dispersion profile data (black dots with error bars). Overplotted is the velocity dispersion profile from 100 randomly selected model parameters sampled by the MCMC (red lines).
    }
    \label{fig:MODEL1SAMPLE}
\end{figure}
\subsection{Dark Matter Density profiles}
\label{subsection:DMdensity}

Dark matter halos in ULDM are well described by a Navarro–Frenk–White \citep[NFW][]{navarro1997} density profile at large radii, $\rho_{\rm NFW}$, and a `soliton core' density profile, $\rho_{\rm DM,s}$, at small radii \citep{schive2014}. The NFW profile is given by: 
\begin{equation}
    \rho_{\rm NFW}(r) = \frac{\rho_{0}}{\frac{r}{r_{\rm s}}\left(1 + \frac{r}{r_{\rm s}}\right)^{2}},
\end{equation}
where $\rho_{0}$ is the characteristic density and $r_{\rm s}$ is the scale radius. The cumulative mass of the NFW profile is given by:
\begin{eqnarray}
  M_{\rm NFW}(<r) & = & \int^r_{0} 4 \pi r'^{2} \rho_{\rm NFW}(r') dr' \nonumber \\
    &= & 4\pi \rho_0 r_{\rm s}^{3} \left[\ln\left(\frac{r_{\rm s}+r}{r_{\rm s}}\right)+\frac{r_{\rm s}}{r_{\rm s}+r}-1\right].
\end{eqnarray}
\citet{schive2014} suggested that the density profile of the soliton core obeys the following equation \citep[e.g.][]{safarzadeh2020}: 
\begin{equation}
    \rho_{\rm DM,s}(r) = \frac{1.9 \{10 [m_{\rm DM}/(10^{-22}~{\rm eV})]\}^{-2}r^{-4}_{\rm c}}{[1 + 9.1 \times 10^{-2}(r/r_{\rm c})^{2}]^{8}}10^{9}~{\rm M_{\odot}~kpc^{-3}},
    \label{Eq:Solitondenisty}
\end{equation}
\begin{equation}
    r_{\rm c} \approx 1.6 [m_{\rm DM}/(10^{-22}~{\rm eV})]^{-1}\Big(\frac{M_{\rm h}}{10^{9}\rm M_{\odot}}\Big)^{-1/3}\rm kpc,
    \label{Eq:rccore}
\end{equation}
where $M_{\rm h}$ is the virial mass of the halo \cite[]{schive2014}. These relations lead to a soliton core mass of:
\begin{equation}
    M_{\rm c} \approx \frac{1}{4}M_{\rm h}^{1/3}(4.4 \times 10^{7}[m_{\rm DM}/(10^{-22}~{\rm eV})]^{-3/2})^{2/3},
    \label{Eq:Solitonmass}
\end{equation}
where $M_{\rm c} \equiv M(<r_{\rm c})$ gives the central core mass \citep[see also][]{safarzadeh2020}.

The total cumulative dark matter mass is, therefore, given by:
\begin{equation}
    M_{\rm DM}(<r) = M_{\rm NFW}(<r)+\int^r_0 4 \pi r'^2 \rho_{\rm DM,s} (r') dr',
\end{equation}
where $\rho_{\rm DM,s}$ is the soliton core density profile of equation (\ref{Eq:Solitondenisty}). We adopt a total mass of the Milky Way of $M_{\rm h}=1.4\times10^{12}~{\rm M_{\odot}}$, with $\rho_0=0.00854$~M$_{\sun}$~pc$^{-3}$ and $r_{\rm s}=19.6$~kpc, obtained from \citet{mcmillan2017}. Once these parameters are fixed, the only free parameter is $m_{\rm DM}$ which controls the shape of the soliton core. As mentioned above, the NFW profile provides a negligible contribution to the total mass within 3~pc, and therefore our analysis is insensitive to  $\rho_{\rm 0}$ or $r_{\rm s}$. However, $M_{\rm h}$ contributes to the soliton core radius and therefore density profile, and it scales as $\rho_{\rm DM,s}\propto M_{\rm h}^{4/3}$ within the core radius. Hence, a larger Milky Way mass produces a denser soliton core, and a larger mass range of the ULDM can, therefore, contribute to the mass within the NSC region -- i.e. a larger mass range of the ULDM can be constrained by the NSC data. 
In fact, the total mass of the Milky Way is still in debate \citep[e.g.][]{erkal2020}. 
Recently, \citet{2021Vasiliev} claims that the virial mass of the Milky Way is as small as $9\times10^{11}$~M$_{\sun}$. In Appendix~\ref{sec:appendixB}, we show the results with $M_{\rm h}=9\times10^{11}$~M$_{\sun}$, and demonstrate that our results are not sensitive to $M_{\rm h}$ as long as it is within the current expected range of $M_{\rm h}$. 
\subsection{Fitting Methodology}
\label{subsection:fitting}
We fit the measured line-of-sight velocity dispersion data in Fig.~\ref{fig:MODEL1SAMPLE} with equation~(\ref{eq:vdisplos}) with our fitting parameters of $m_{\rm DM}$, $\rho_{\rm b,NSC}$, $\gamma$ and $m_{\rm BH}$. We include the SMBH mass of $m_{\rm BH}$ as a fitting parameter, because the SMBH mass is dominant at radii $r\leq1$~pc. We use Bayesian statistics to obtain the marginalised probability distribution function for these parameters, $\theta_{\rm m}=(m_{\rm DM}, \rho_{\rm b,NSC}, \gamma, m_{\rm BH})$:
\begin{equation}
    P(\theta_{\rm m} | D) = \mathcal{L}(D|\theta_{\rm m}) \times prior,
\end{equation}
where $D$ is the data, i.e. the line-of sight velocity dispersion in different radial bins (Fig.~\ref{fig:MODEL1SAMPLE}). 

To obtain $P(\theta_{\rm m} | D)$, we run a Markov Chain Monte Carlo (MCMC) fit, with a likelihood function given by:
\begin{equation}
    \mathcal{L}(D| \theta_{\rm m})=\prod_i^{N_{\rm D}} \frac{1}{\sqrt{2 \pi \sigma_{\rm err,i}^{2}}} 
 \exp\left( - \frac{(\sigma_{\rm m}(R_{i}, \theta_{m})-\sigma_{\rm obs,i})^2}{2 \sigma_{\rm err,i}^{2}} \right),
\end{equation}
where $\sigma_{\rm obs,i}$ is the observed line-of-sight velocity dispersion data at $R_i$, $\sigma_{\rm err,i}$ is the measurement error on each bin, $N_{\rm D}$ is the number of the data points, and $\sigma_{\rm m}(R_{i}, \theta_{m})$ is the model line-of-sight velocity dispersion at $R_{i}$ (with parameters $\theta_{\rm m}$). 

We use $\log(\rho_{\rm b,NSC})$ and $\log(m_{\rm DM})$ as our fitting parameters with flat priors of $3 < \log[\rho_{\rm b,NSC}{\rm (M_{\sun}~pc^{-3})}]  < 7$ and $-23 < \log[m_{\rm DM}{\rm (eV)}] < -16$, since we find that the likelihood changes more smoothly in $\log(\rho_{\rm b,NSC})$ and $\log(m_{\rm DM})$. The range of  $\log[\rho_{\rm b,NSC}{\rm (M_{\sun}~pc^{-3})}]$ is chosen as above, because outside of this range is unrealistic from the NSC photometric observations \citep[e.g.][]{Schoedel+Feldemeier+Kunneriath+14}. Since $\gamma$ and $m_{\rm BH}$ are well-constrained by the other observations, as described above, we adopt Gaussian priors for these two parameters. The Gaussian prior for $\gamma$ has a mean and dispersion of 1.29 and 0.05, respectively. The mean and dispersion for the Gaussian prior on $m_{\rm BH}$ are set to be $4.26\times10^6$~M$_{\sun}$ and $0.012\times10^6$~M$_{\sun}$, respectively.

In Fig.~\ref{fig:densityprofiles}, we show the NSC density profile (higher mass solution inferred in Section~\ref{sec:results}) and the ULDM dark matter density profile with $m_{\rm DM}=10^{-23.0}$~eV, $10^{-21.0}$~eV, $10^{-20.0}$~eV, $10^{-19.0}$~eV, $10^{-18.0}$~eV, $10^{-16.0}$~eV and the NFW dark matter density profile. Fig.~\ref{fig:densityprofiles} shows that a soliton core with higher ULDM mass has a higher density at the centre, but a smaller core size. Consequently, within the radial range where we focus in this paper, i.e. $0.1<r<3$~pc, only the ULDM soliton core with a mass range of about $10^{-20.0}<m_{\rm DM}<10^{-19.0}$~eV becomes important, compared to the NSC. In other words, the NSC kinematics in this radial range has the potential to constrain the existence of $10^{-20.0}<m_{\rm DM}<10^{-19.0}$~eV ULDM, as discussed in \citet{bar2018}. Fig.~\ref{fig:densityprofiles} also shows that the soliton core with $m_{\rm DM}<10^{-23.0}$~eV or $m_{\rm DM}>10^{-16.0}$~eV has negligible density within $0.1<r<3$~pc as compared to the expected NSC density. Hence, we consider that our prior range on $\log(m_{\rm DM})$ is large enough to capture the region we hope to constrain. 

We use {$\tt$ emcee} \citep{foreman2013} for our MCMC sampler, with 32 walkers and 4000 chains per walker. We discard the first 1000 chains as our `burn-in'. We confirm that after 1000 steps the MCMC results are stable.

\begin{figure*}
    \centering
    \includegraphics[width = \textwidth]{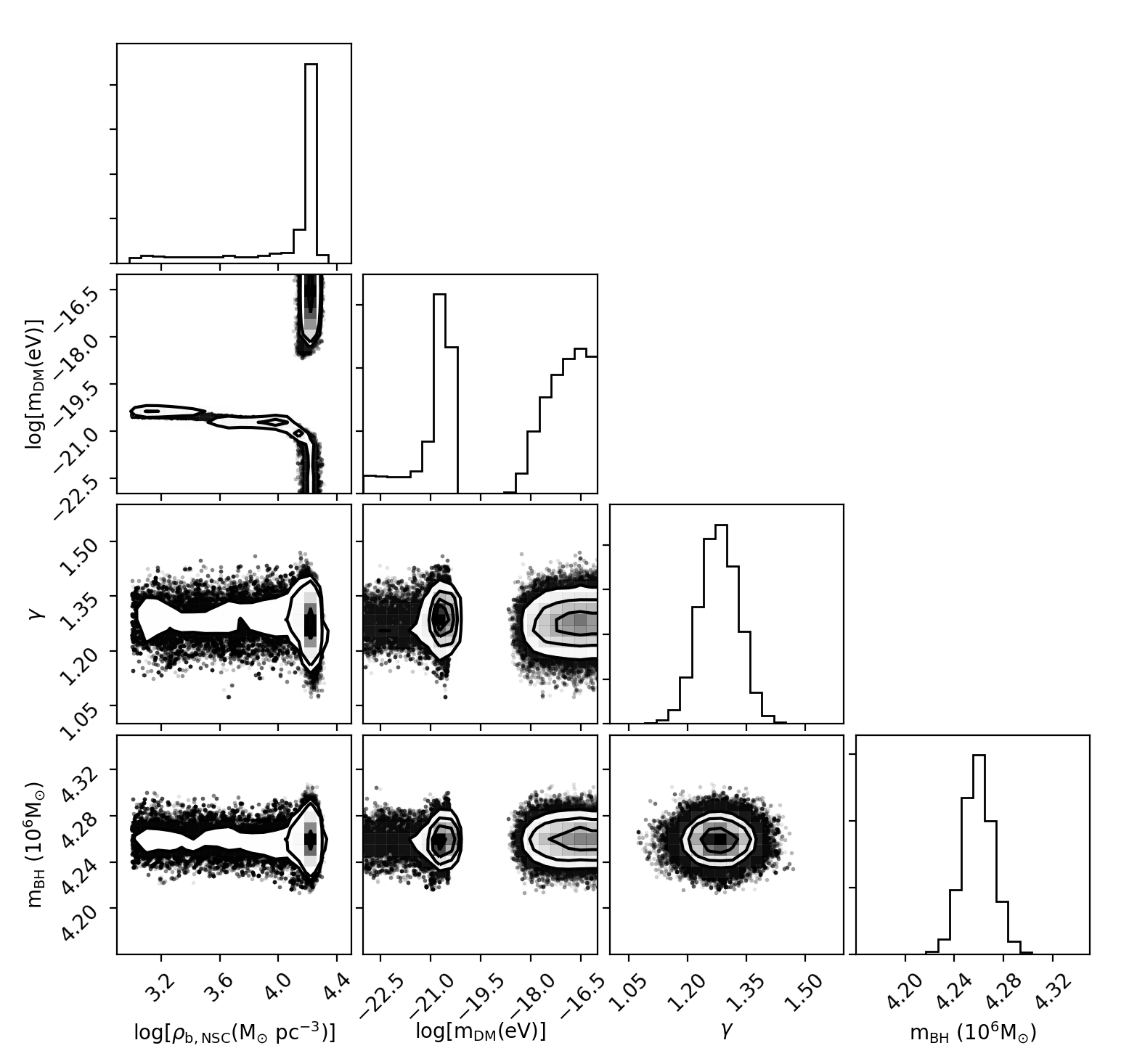}
    \caption{Marginalised posterior probability distribution of the model parameters of $\log(\rho_{\rm b,NSC})$, $\log(m_{\rm DM})$, $\gamma$ and $m_{\rm BH}$ obtained by MCMC fitting to the observed velocity dispersion from the line-of-sight velocity data in \citet{fritz2016}.}
    \label{fig:MODEL1CON}
\end{figure*}

\section{Results}
\label{sec:results} 

Fig.~\ref{fig:MODEL1SAMPLE} shows our modelled line-of-sight velocity dispersion profiles (eq.~\ref{eq:vdisplos}) for 100 random parameter values sampled from our MCMC chains, as compared to the observed velocity dispersion data. Notice that there is a good agreement between the sampled line-of-sight velocity dispersion profiles and the observational data.

Fig.~\ref{fig:MODEL1CON} shows the marginalised posterior probability distribution of our fitting parameters of $\log(\rho_{\rm b,NSC})$, $\log(m_{\rm DM})$, $\gamma$ and $m_{\rm BH}$. Notice that $\gamma$ and $m_{\rm BH}$ are well constrained. We compute the mean and standard deviation of the posterior probability distributions of these parameters and obtain the best-fitting parameter values and 1$\sigma$ uncertainties of $\gamma= 1.28 \pm 0.04$ and $m_{\rm BH} = (4.26\pm 0.01)\times10^{6}$~M$_{\sun}$. Our results show that the best-fitting values of $\gamma$ and $m_{\rm BH}$ are consistent with our priors, i.e. the observed inner slope of the NSC measured by \citet{gallego2018} and the black hole mass measured by \citet{abuter2020detection}.

\begin{figure}
    \centering
    \includegraphics[width = \columnwidth]{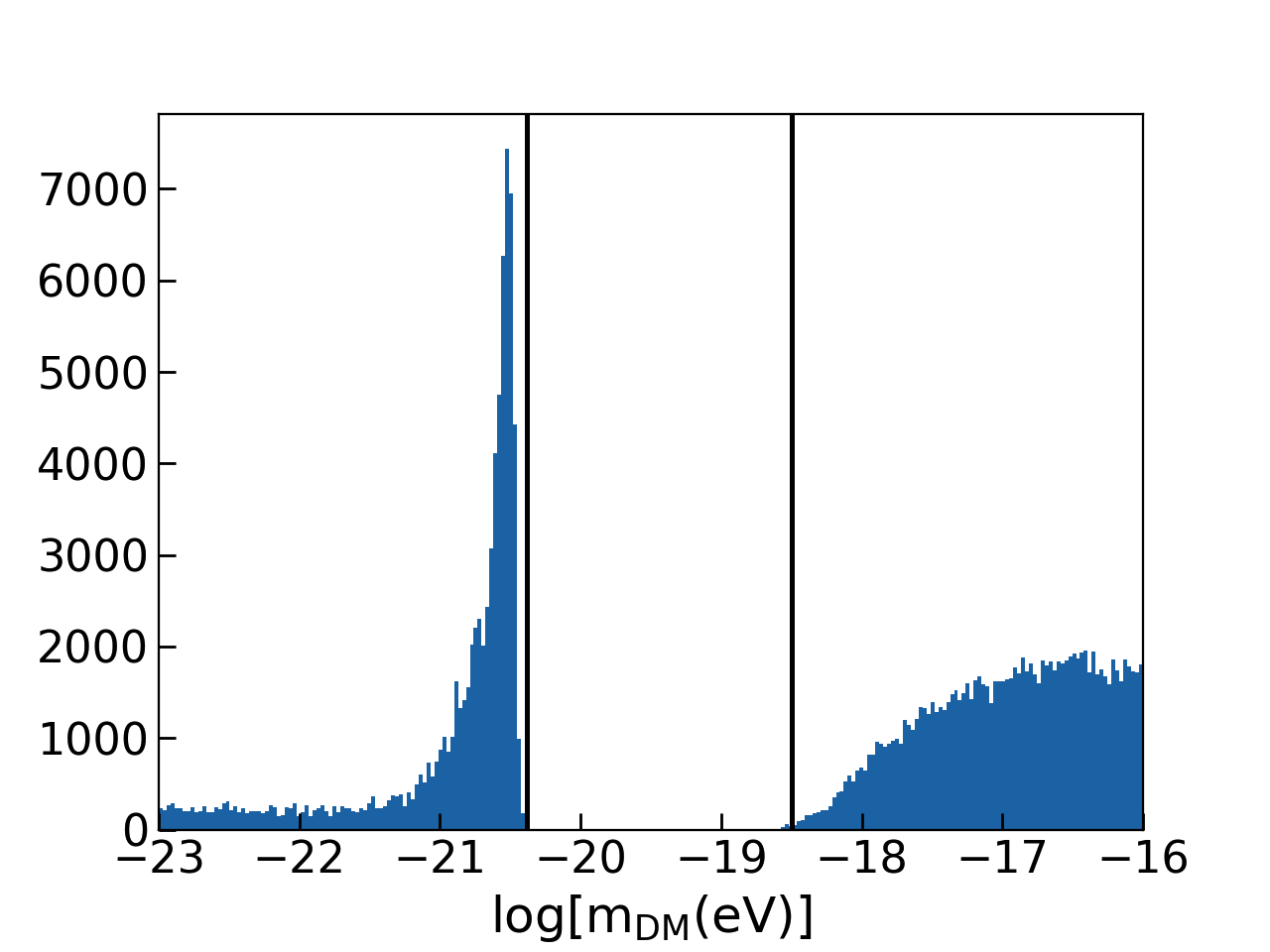}
    \caption{Marginalised posterior probability distribution of the model parameter $\log(m_{\rm DM})$ from Fig.~\ref{fig:MODEL1CON}, but with finer bins. The solid black lines demark $\log[m_{\rm DM}{\rm (eV)}]=-20.4$ and $-18.5$.}
    \label{fig:MODEL1Hist}
\end{figure}

\begin{figure}
    \centering
    \includegraphics[width = \columnwidth]{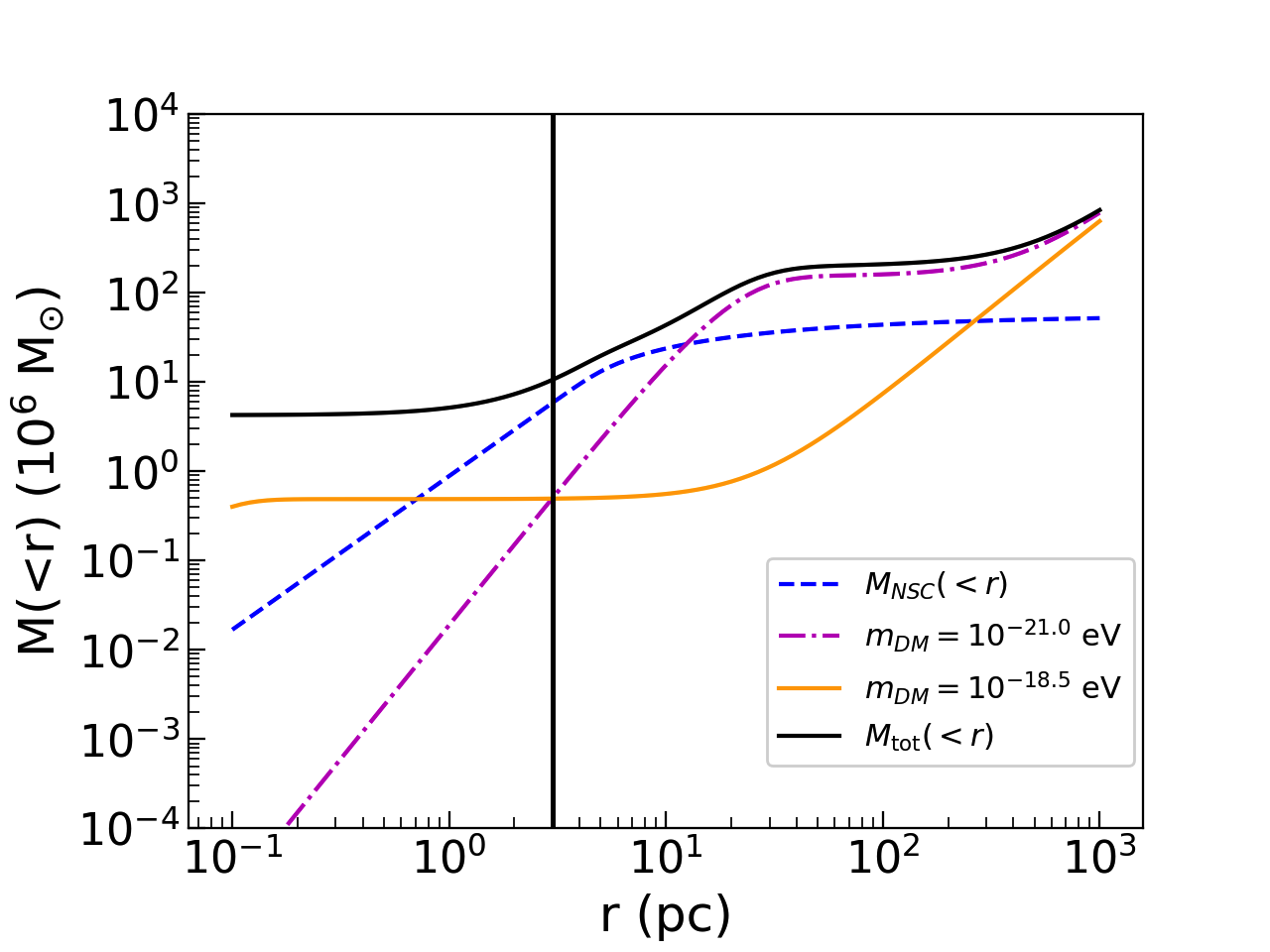}
    \caption{The cumulative mass profile, $M_{\rm tot}(<r)$, for the total (black solid line), NSC (blue solid line) and dark matter with a ULDM particle mass of $10^{-18.5}$~eV and $10^{-21.0}$~eV
    (orange solid and magenta dot-dashed lines, respectively). The solid vertical black line shows $r=3$~pc. The NSC mass profile is computed with $\log[\rho_{\rm b,NSC}{\rm (M_{\sun}~pc^{-3})}]=4.21$. The total mass is computed for the case of the ULDM mass of $m_{\rm DM}=10^{-21.0}$~eV, including the SMBH.
    }
    \label{fig:MODEL1CM2}
\end{figure}

\begin{figure}
    \centering
    \includegraphics[width = \columnwidth]{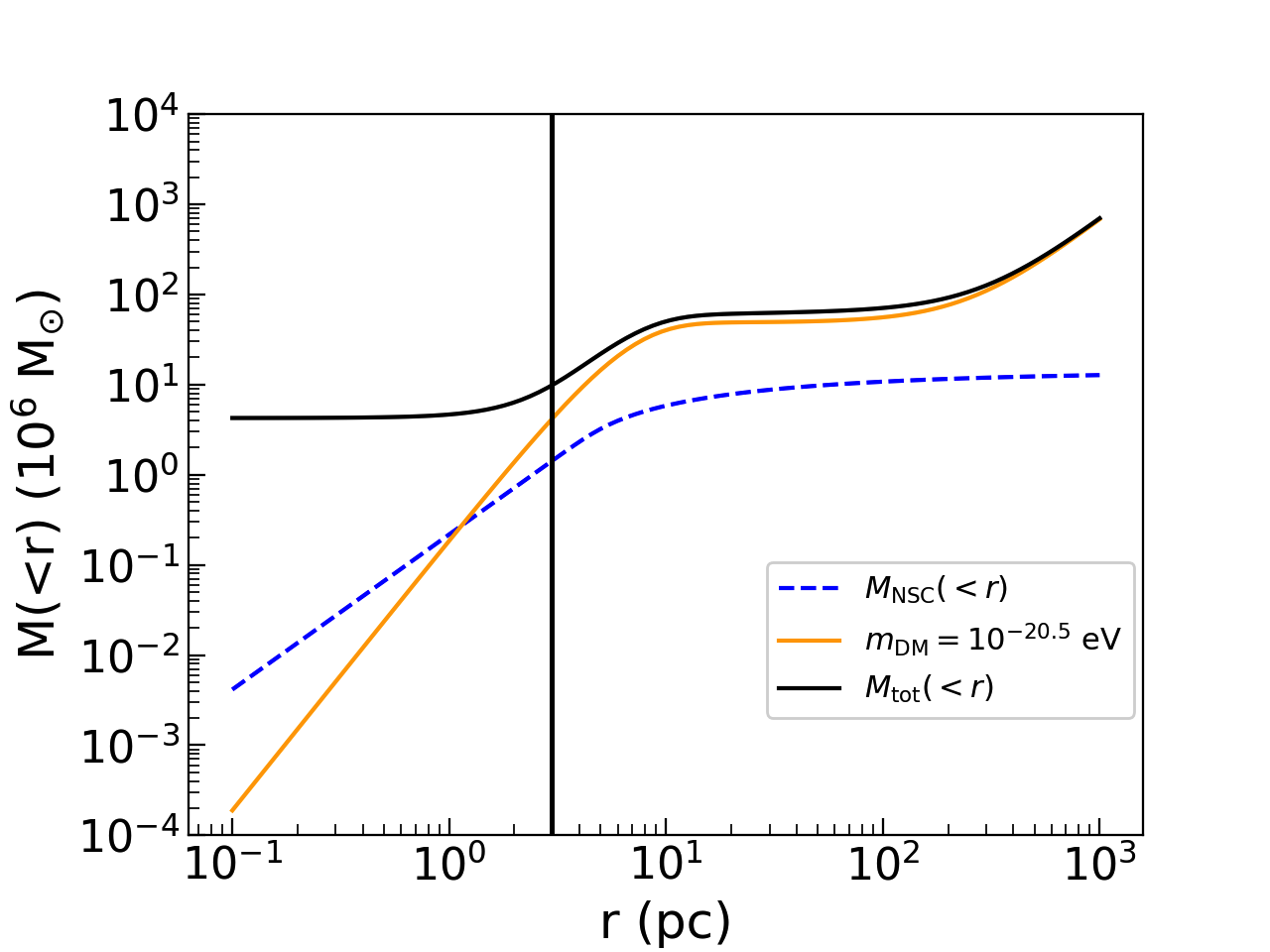}
    \caption{The cumulative mass profile, $M_{\rm tot}(<r)$, for the total (black), NSC (blue) and dark matter mass with the particle mass of $10^{-20.5}$~eV (orange). The solid vertical black line shows $r=3$~pc. The total mass is computed for the case of the ULDM mass of $m_{\rm DM}=10^{-20.5}$~eV, including the SMBH.}
    \label{fig:MODEL1CM}
\end{figure}

\begin{figure}
    \centering
    \includegraphics[width = \columnwidth]{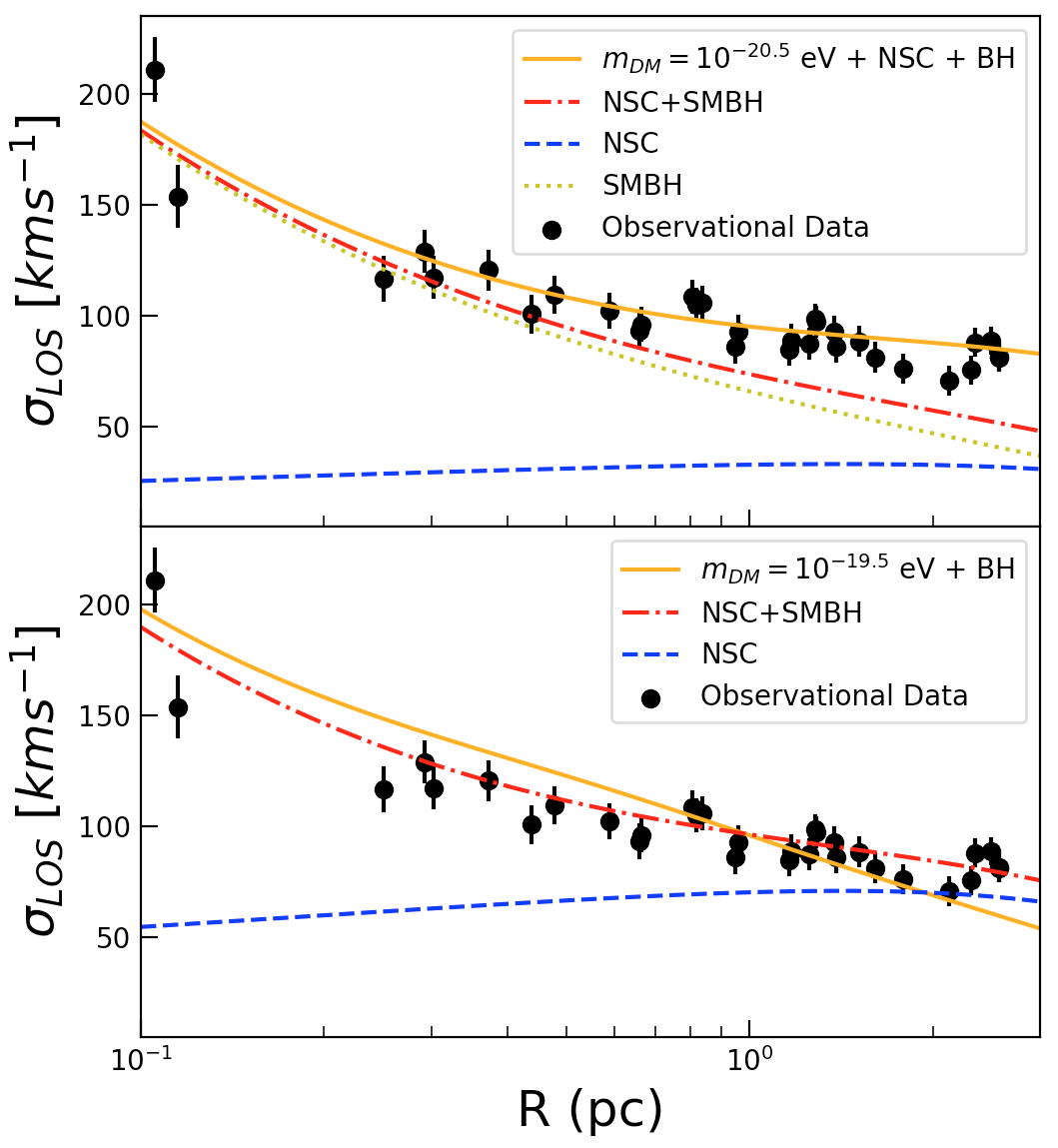}
        \caption{Upper panel: Observed line-of-sight velocity dispersion as a function of the projected radius (black dots with error bars). Orange solid/red dot-dashed/yellow dotted/blue dashed line indicates the velocity dispersion profile expected from the combination of the soliton core with $m_{\rm DM}=10^{-20.5}$~eV, NSC and SMBH/NSC and SMBH/SMBH only/NSC only. NSC contribution is computed with $\log[\rho_{\rm b,NSC} ({\rm M_{\odot}~pc^{-3}})]=3.60$. Lower panel: Same as the upper panel, but the soliton core with $m_{\rm DM}=10^{-19.5}$~eV and $\log[\rho_{\rm b,NSC}{\rm (M_{\sun}~pc^{-3})}]= 4.21$ are used for the soliton core and NSC contributions.}
    \label{fig:allVDprofiles}
\end{figure}

Fig.~\ref{fig:MODEL1Hist} shows a close-up view of the marginalised probability distribution of $\log(m_{\rm DM})$ with a histogram with a smaller bin size, where we can see two interesting results. First is the gap of the posterior probability distribution of $\log(m_{\rm DM})$ around the range of $-20.4\lesssim \log[m_{\rm DM}{\rm (eV)}] \lesssim -18.5$, which is highlighted by the black vertical lines of $\log[m_{\rm DM}{\rm (eV)}]=-20.4$ and $-18.5$ in Fig.~\ref{fig:MODEL1Hist}. 
This result indicates that the observational data reject the ULDM particle mass between about  $10^{-20.4}$~eV and 
$10^{-18.5}$~eV. 

Note that the upper and lower limits of $\log(m_{\rm DM})$ in Fig.~\ref{fig:MODEL1Hist} come from the upper and lower limit of the flat prior we imposed. The roughly flat probability distributions at higher than about $-18.5$ and lower than about $-21.0$ mean that the observational data cannot distinguish the difference in the ULDM particle mass in these ranges. 
Fig.~\ref{fig:MODEL1CM2} shows the cumulative mass profiles of the NSC, dark matter and the total mass as a function of the Galactocentric 3D radius. For the NSC profile, we take  $\log[\rho_{\rm b,NSC}{\rm (M_{\sun}~pc^{-3})}]= 4.21$, which is the mean $\log(\rho_{\rm b,NSC})$ of our MCMC samples with $\log[m_{\rm DM}{\rm (eV)}] > -18.0$ or $\log[m_{\rm DM}{\rm (eV)}] < -21.0$.
This leads to a NSC mass within $r=3$~pc of about $5.03\times10^{6}$~M$_{\odot}$, which is larger than the value of about $3.965\times10^{6}$~M$_{\odot}$ measured by \citet{fritz2016} within 75~arcsec ($r\sim3$~pc). This is likely due to different density profiles we are using. 
For example, \citet{fritz2016} uses a lower $\gamma$ value of $\gamma$ = 0.81. We tested our results with a Gaussian prior for $\gamma$ with the mean value of 0.81 and we confirmed that the NSC mass within 3~pc reduced to $3.91\times10^{6}$~M$_{\odot}$, which is similar to the measured value by \citet{fritz2016}. 

Fig.~\ref{fig:MODEL1CM2} also shows the cumulative mass profile of the ULDM with $m_{\rm DM}=10^{-21.0}$~eV and $m_{\rm DM}=10^{-18.5}$~eV, where both cumulative masses reach about $4.4\times10^{5}$~M$_{\odot}$ at 3\,pc. These two ULDM soliton cores are much smaller than both the NSC mass within the same radius and the SMBH mass.  
Because the size of the soliton core increases with decreasing particle mass of the ULDM (eq.~\ref{Eq:rccore}), the soliton core mass within $r<3$~pc decreases with the decreasing ULDM particle mass. Consequently, the ULDM particles mass with $m_{\rm DM}<10^{-21.0}$~eV does not affect the velocity dispersion of the NSC. This explains the equally accepted probability distribution of $m_{\rm DM}<10^{-21.0}$~eV in Fig.~\ref{fig:MODEL1Hist}. On the other hand, the ULDM particles mass with $m_{\rm DM}>10^{-18.5}$~eV leads to too small of a soliton core to affect the stellar dynamics in the central region. This explains the equally accepted probability distribution at $m_{\rm DM}>10^{-18.5}$~eV.
Hence, if the ULDM particle mass is larger than $m_{\rm DM}=10^{-18.5}$~eV or smaller than $m_{\rm DM}=10^{-21.0}$~eV, our current data of the NSC stellar dynamics cannot find or reject their existence.
\begin{figure}
    \centering
    \includegraphics[width = \columnwidth]{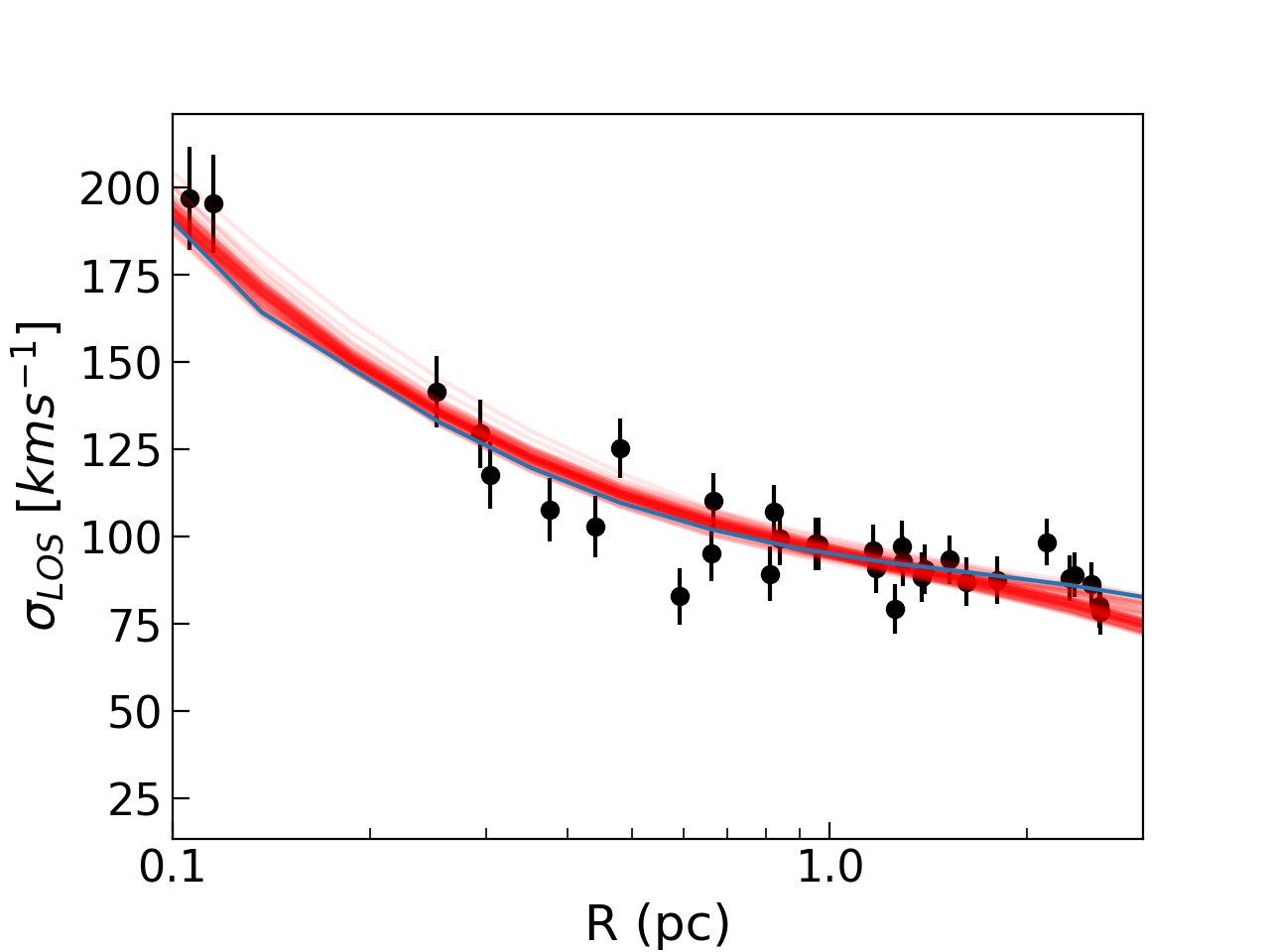}
    \caption{The mock line-of-sight velocity dispersion profile data of model A (black dots with error bars) overplotted with the velocity dispersion profile from 100 randomly selected model sampled by the MCMC (red lines) and the true velocity dispersion profile (blue line).}
    \label{fig:MockModel1VD}
\end{figure}
\begin{table*}
\caption{Model parameters of the mock data.}
\centering
 \begin{tabular}{c c c c c} 
 \hline
 Model name & $\log[m_{\rm DM}{\rm (eV)}]$ & $\log[\rho_{\rm b,NSC}{\rm (M_{\sun}~pc^{-3})}]$ & $\gamma$  & $m_{\rm BH}$ \\
 [0.5ex] 
 \hline
 A & $-20.5$ & $3.60$ & 1.29 
  & 4.26 \\ 
 B & $\rm -19.5$ & $4.50$ & 1.29 & 4.26 \\
 C & $\rm -23.0$ & $4.21$ & 1.29  & 4.26 \\
 \hline
\end{tabular}
\label{table:MockParam}
\end{table*}
\begin{figure*}
    \centering
    \includegraphics[width = \textwidth]{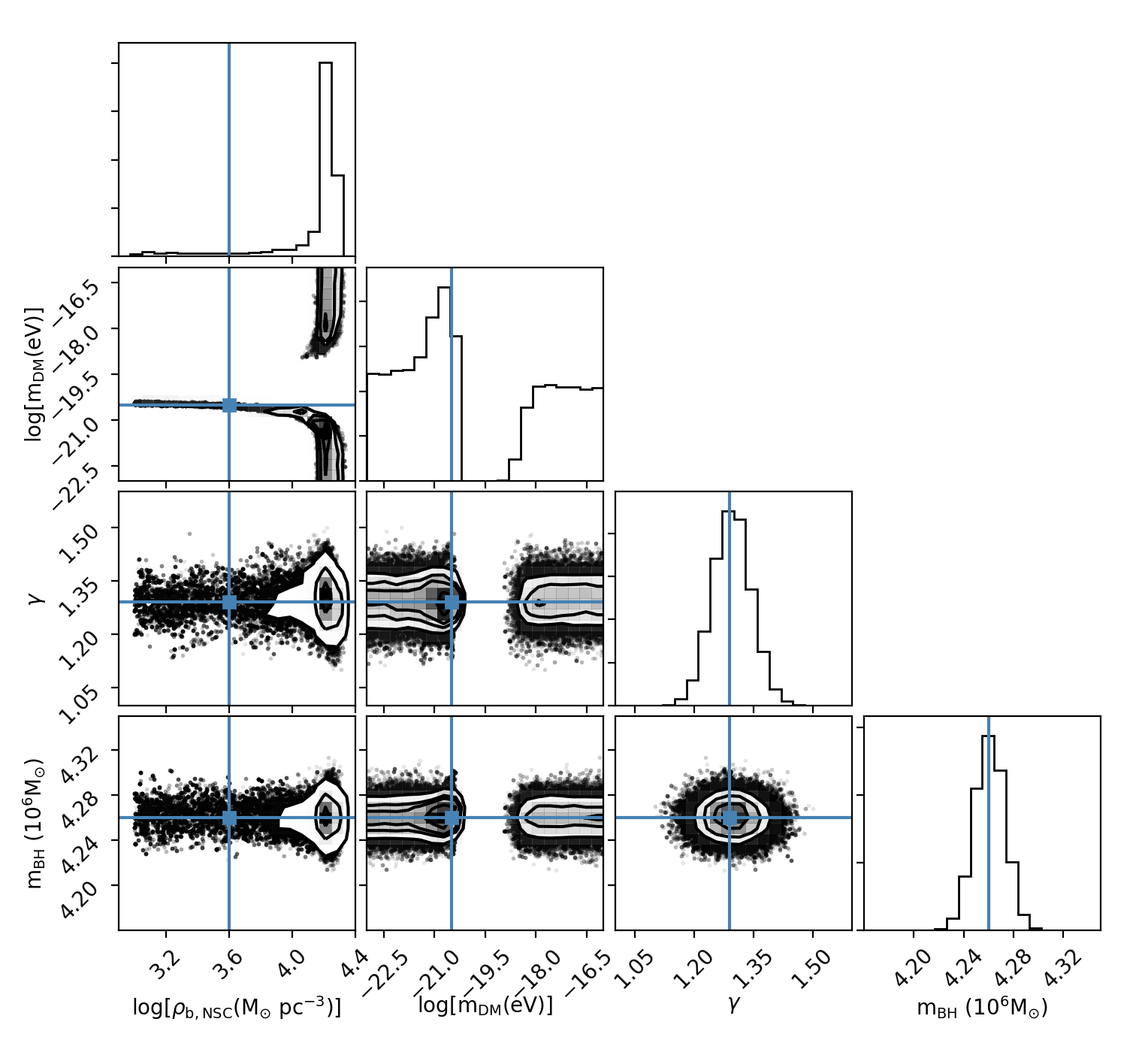}
    \caption{Marginalised posterior probability distribution of the model parameters of $\log(\rho_{\rm b,NSC})$, $\log(m_{\rm DM})$, $\gamma$ and $m_{\rm BH}$ obtained by the MCMC fit to the velocity dispersion data of model~A. The cyan line with cyan solid square shows the true values of the parameters.}
    \label{fig:MockModel1Cont}
\end{figure*}

The second striking result of Fig.~\ref{fig:MODEL1Hist} is the peak around $\log[m_{\rm DM}{\rm (eV)}]=-20.5$. At first sight, this appears to statistically favour a soliton core due to ULDM with a mass of $m_{\rm DM}=10^{-20.5}$~eV. Fig.~\ref{fig:MODEL1CM} shows the cumulative mass profile for the total mass, dark matter halo mass including the soliton core with $m_{\rm DM} = 10^{-20.5}$~eV and the NSC mass with $\log[\rho_{\rm b,NSC}{\rm (M_{\sun}~pc^{-3})}] = 3.60$, which is the mean of $\log(\rho_{\rm b,NSC})$ of the MCMC sample with $-21.0<\log[m_{\rm DM}{\rm (eV)}]<-20.4$. Fig.~\ref{fig:MODEL1CM} shows that the NSC mass is smaller than that in Fig.~\ref{fig:MODEL1CM2}, and the $m_{\rm DM}=10^{-20.5}$~eV soliton core has a suitable size to compensate the deficit of the mass within $r<3$~pc. The upper panel of Fig.~\ref{fig:allVDprofiles} also shows that the additional mass from the $m_{\rm DM}=10^{-20.5}$~eV soliton core helps to increase the velocity dispersion at an outer radius ($r>0.5$~pc) to match with the observational data more than the expected velocity dispersion from the NSC and SMBH only. 

Consequently, the NSC mass within 3~pc is about $1.25 \times 10^{6}$~$\rm M_{\odot}$, which is significantly smaller than the aforementioned NSC mass measured by \citet{fritz2016}. The cumulative mass of the NSC in Fig.~\ref{fig:MODEL1CM} is also much smaller than the NSC mass of $(2.1\pm0.7)\times10^7$~M$_{\sun}$ within about 8.4~pc, as measured in \citet{Feldmeier-Krause+Zhu+Neumayer+17}. Although these studies use dynamical models that assume that the NSC is the dominant source of the central gravitational potential, the photometric observations of \citet{Schoedel+Feldemeier+Kunneriath+14} also suggest a total NSC mass of $(2.5\pm0.4)\times10^7$~M$_{\sun}$, assuming a constant mass to light ratio. Hence, it is unlikely that the NSC mass is as small as the case of Fig.~\ref{fig:MODEL1CM}. Thus, the peak of $m_{\rm DM}=10^{-20.5}$~eV is not likely to be a viable solution. Still, it is difficult to measure the mass to light ratio precisely, and there could be some systematic biases in these previous measurements. Hence, we consider that we cannot (yet) reject the existence of the $m_{\rm DM}=10^{-20.5}$~eV soliton core.

The constraining power of the observed velocity dispersion data to reject the ULDM mass between about  $10^{-20.4}$~eV and 
$10^{-18.5}$~eV in  Fig.~\ref{fig:MODEL1Hist} can
be demonstrated in the lower panel of Fig.~\ref{fig:allVDprofiles}. The lower panel of Fig.~\ref{fig:allVDprofiles} shows that the velocity dispersion profile expected from the $m_{\rm DM}=10^{-19.5}$~eV soliton core and SMBH even without NSC (orange line) is systematically higher than the observational data within $r=1$~pc. Hence, the data can reject the soliton core with the ULDM mass around $10^{-19.5}$~eV. On the other hand, the velocity dispersion profile expected from the SMBH and NSC with $\log[\rho_{\rm b,SMC}{\rm (M_{\odot} pc^{-3})}] = 4.21$, i.e. without any soliton core (red doted-dashed line), agrees well with the observational data. Hence, NSC and SMBH are enough to describe the observed stellar kinematics. 

\section{Mock Data Validation}
\label{sec:Mock}
In Section~\ref{sec:results}, we found a gap in the probability distribution function of ULDM masses that rejects a ULDM particle in the mass range $-20.4\lesssim \log[m_{\rm DM}{\rm (eV)}] \lesssim -18.5$. We also found a peak in the probability distribution around $\log[m_{\rm DM}{\rm (eV)}]=-20.5$ that we argued owed to a degeneracy between $\rho_{\rm b,NSC}$ and $m_{\rm DM}$.

To test the voracity of above results, we construct mock velocity dispersion data similar to the observational data, using the same model as in Section~\ref{sec:method}. We then fit the data as in Section~\ref{sec:results}. We adopt the same parameters for the SMBH, NSC and dark matter model as in Section~\ref{sec:method}.

We construct three different models with different values of $\rho_{\rm b,NSC}$ and $m_{\rm DM}$, as shown in Table~\ref{table:MockParam}. We then generate the mock velocity dispersion profile data for each model by solving equation~(\ref{eq:vdisplos}) for 32 bins spaced out in exactly the same way as for the observational data. We then add a random displacement to the velocity dispersion of each bin, within the measurement error of each bin, taken to be the same as for the observational data.

We use the same fitting methodology with the same priors, as described in Section \ref{subsection:fitting}, except that now the observational data are replaced by mock data for three models, labelled A, B and C (Table \ref{table:MockParam}).

Model A employs $m_{\rm DM}= 10^{-20.5}$~eV and $\log[\rho_{\rm b,NSC}{\rm (M_{\sun}~pc^{-3})}]=3.60$, which is the mean value of our MCMC samples around $m_{\rm DM}= 10^{-20.5}$~eV found in Section~\ref{sec:results}. This model is to test if the probability distribution of $\log[m_{\rm DM}{\rm (eV)}]$ would be similar to what is obtained in Fig.~\ref{fig:MODEL1Hist}, when a soliton core of the $m_{\rm DM}= 10^{-20.5}$~eV UDLM exists.

\begin{figure}
    \centering
    \includegraphics[width = \columnwidth]{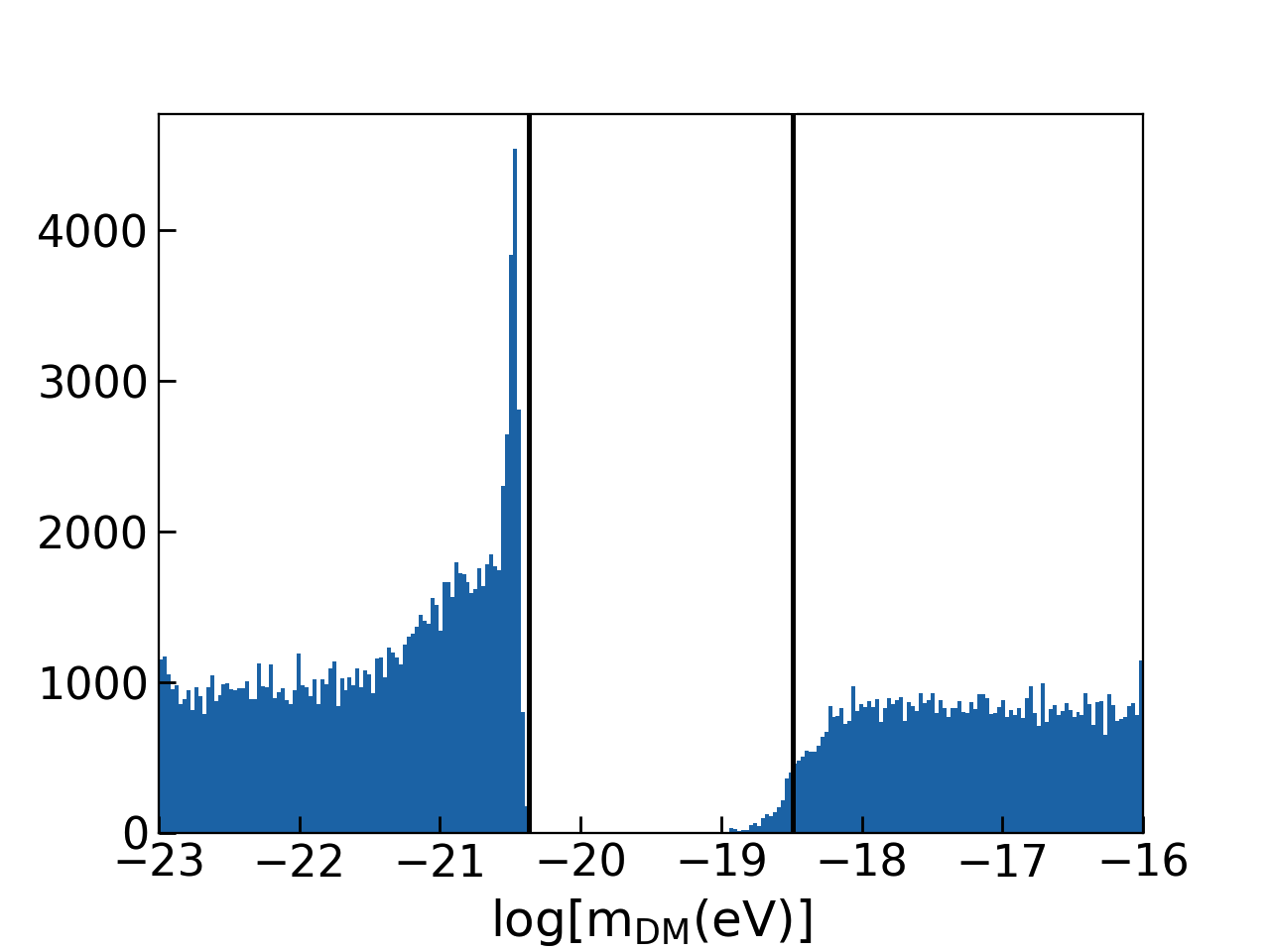}
    \caption{Marginalised posterior probability distribution of the model parameter $\log(m_{\rm DM})$ for model~A from Fig.~\ref{fig:MockModel1Cont}, but with finner bins. The solid black lines demark the range $\log[m_{\rm DM}{\rm (eV)}]=-20.4$ and $-18.5$.}
    \label{fig:MockModel1Hist}
\end{figure}

Fig.~\ref{fig:MockModel1VD} overplots the model line-of-sight velocity dispersion profiles from the 100 random parameter values sampled from the results of MCMC with the mock velocity dispersion data for model~A. Fig.~\ref{fig:MockModel1VD} shows that there is a good agreement between the sampled line-of-sight velocity dispersion profiles and the mock data roughly within the uncertainties of the mock data. Fig.~\ref{fig:MockModel1Cont} shows the marginalised posterior probability distribution of our fitting parameters of $\log(\rho_{\rm b,NSC})$, $\log(m_{\rm DM})$, $\gamma$ and $m_{\rm BH}$ for model~A with the cyan line with the cyan solid square representing the true values of the parameters.

The obtained best-fitting parameter values and 1$\sigma$ uncertainties are $\gamma= 1.29\pm0.05$ and $m_{\rm BH} = (4.26 \pm 0.01)\times10^{6}$~M$_{\sun}$, which are consistent with the true values within our 1$\sigma$ uncertainty regions. Just like the results in Section~\ref{sec:results}, there is a degeneracy between $\log(\rho_{\rm b,NSC})$ and $\log(m_{\rm DM})$. 
In the probability distribution between $\log(\rho_{\rm b,NSC})$ and $\log(m_{\rm DM})$, when $\log[m_{\rm DM}{\rm (eV)}]$ is around the true value of $-20.5$, $\log(\rho_{\rm b,NSC})$ corresponds to $\log[\rho_{\rm b,NSC}{\rm (M_{\sun}~pc^{-3})}] = 3.74\pm0.37$, which is within one sigma of the true value of $\log[\rho_{\rm b,NSC}{\rm (M_{\sun}~pc^{-3})}] = 3.60$.

The close-up plot of the marginalised probability distribution of ${\rm log}(m_{\rm DM})$ is shown in
Fig.~\ref{fig:MockModel1Hist}, and there is a similar peak around about $10^{-20.5}$~eV when compared to Fig.~\ref{fig:MODEL1Hist}. Also, Fig.~\ref{fig:MockModel1Hist} shows the gap between $\sim-20.4 \lesssim \log[m_{\rm DM}~{\rm (eV)}] \lesssim \sim-18.5$, and roughly flat probability distribution at $\log[m_{\rm DM}~{\rm (eV)}]<-21.0$ and $\log[m_{\rm DM}~{\rm (eV)}]>-18.5$, as seen in Fig.~\ref{fig:MockModel1Hist}. This implies that the result in Section~\ref{sec:results} is consistent with the expected result when there is a soliton core with ULDM particle mass around $10^{-20.5}$~eV.

Model B adopts $\log[\rho_{\rm b,NSC}{\rm (M_{\sun}~pc^{-3})}]=4.50$ and $m_{\rm DM} = 10^{-19.5}$~eV, to see if the data are capable of detecting a soliton core with $m_{\rm DM} = 10^{-19.5}$~eV. If it is confirmed, we can be confident that the gap we obtained in Fig.~\ref{fig:MODEL1Hist} in Section \ref{sec:results} is not due to an artificial feature, but rather it is meaningful to reject the existence of a soliton core over this mass range. The choice of this higher $\log[\rho_{\rm b,NSC}{\rm (M_{\sun}~pc^{-3})}]$ compared to models~A and C is to make the NSC more gravitationally dominant, i.e. to make it more challenging to recover the soliton core contribution.

\begin{figure*}
    \centering
    \includegraphics[width = \textwidth]{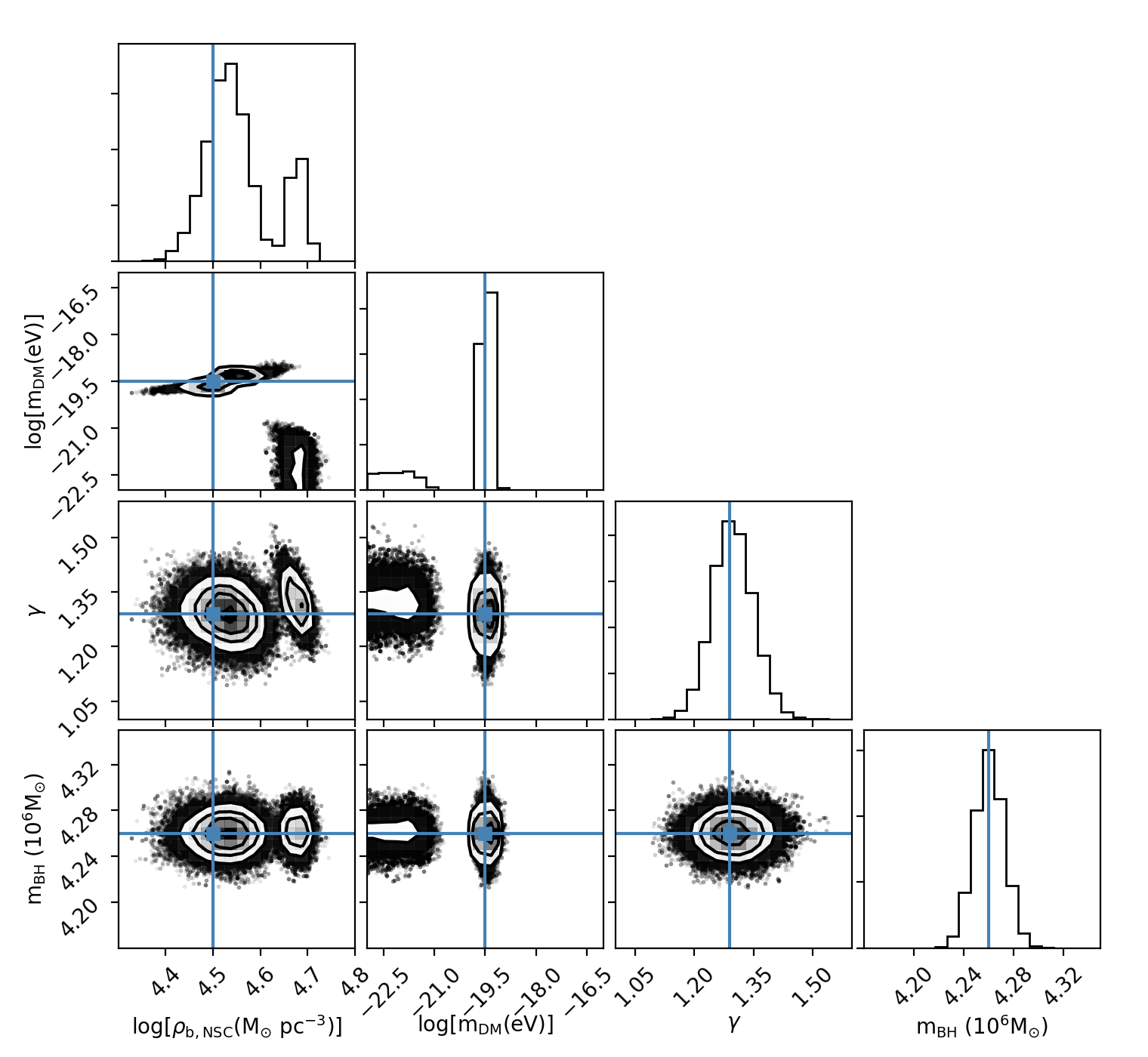}
    \caption{Marginalised posterior probability distribution of the model parameters of $\log(\rho_{\rm b,NSC})$, $\log(m_{\rm DM})$, $\gamma$ and $\rm m_{\rm BH}$ obtained by the MCMC fitting to the velocity dispersion data of model~B. The cyan line with cyan solid square shows the true values of the parameters.}
    \label{fig:MockModel2Cont}
\end{figure*}

Although not shown for brevity, we confirm that there is a good agreement between the sampled line-of-sight velocity dispersion profiles and the mock data of model~B within the uncertainties of the mock data. Fig.~\ref{fig:MockModel2Cont} shows the marginalised posterior probability distribution of our fitting parameters for model~B with the cyan line with the cyan solid square representing the true values of the parameters. The best fitting values and the respective uncertainties of the parameters are $\log[\rho_{\rm b,NSC}{\rm (M_{\sun}~pc^{-3})}]= 4.56 \pm 0.07$, $\log[m_{\rm DM}{\rm (eV)}] = -19.51 \pm 1.09$, $\gamma= 1.30 \pm 0.05$ and $m_{\rm BH} = (4.26 \pm 0.01)\times10^{6}$~M$_{\sun}$, which are consistent with the true value within our 1$\sigma$ uncertainty regions. This demonstrates that our MCMC fitting can recover the true parameter values well, especially the ULDM particle mass, which is the main focus of this paper. This means that the current observational data are good enough to identify a soliton core of $m_{\rm DM}=10^{-19.5}$~eV, if it exists.

\begin{figure*}
    \centering
    \includegraphics[width = \textwidth]{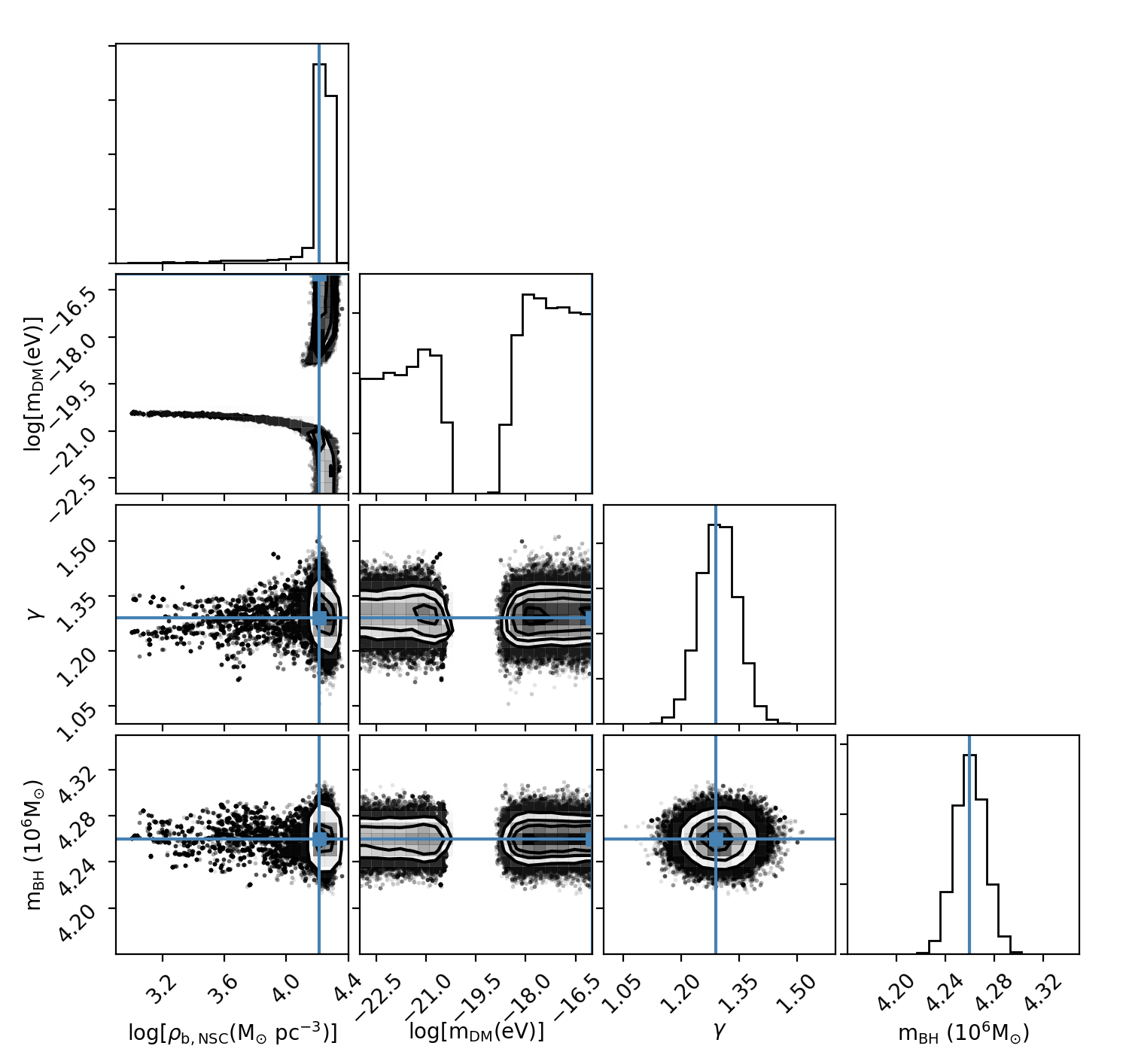}
    \caption{Marginalised posterior probability distribution of the model parameters of $\log(\rho_{\rm b,NSC})$, $\log(m_{\rm DM})$, $\gamma$ and $\rm m_{\rm BH}$ obtained by the MCMC fit to the velocity dispersion data of model~C. The cyan line with cyan solid square shows the true values of the parameters.}
    \label{fig:MockModel3Cont}
\end{figure*}

\begin{figure}
    \centering
    \includegraphics[width = \columnwidth]{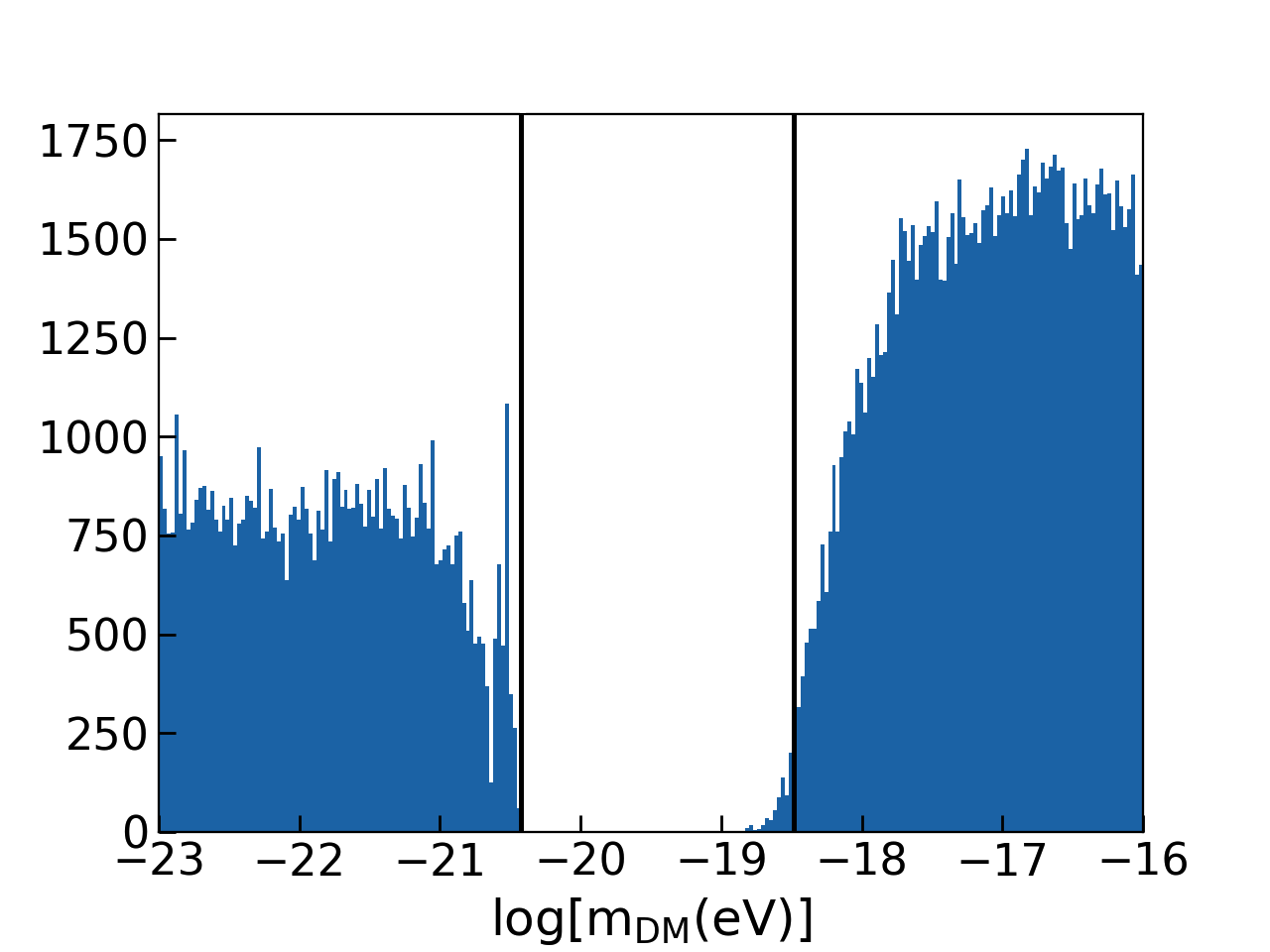}
    \caption{Marginalised posterior probability distribution of the model parameter $\log(m_{\rm DM})$ for model~C from Fig.~\ref{fig:MockModel3Cont}, but with finer bins. The solid black lines demark $\log[m_{\rm DM}{\rm (eV)}]=-20.4$ and $-18.5$.}
    \label{fig:MockModel3Hist}
\end{figure}

Model C employs $m_{\rm DM} = 10^{-23.0}$~eV. As we discussed in Section~\ref{sec:results}, this particle mass of ULDM produces a negligible soliton core mass compared to the SMBH and NSC mass (see also Fig.~\ref{fig:densityprofiles}), i.e. mimicking the case of no detectable soliton core. Hence, this model is designed to test what our MCMC fitting results will look like if there is no soliton core. Model C adopts $\log[\rho_{\rm b,NSC}{\rm (M_{\sun}~pc^{-3})}] = 4.21$, which is found to be the best fitting parameter in Section~\ref{sec:results}, when the soliton core is negligible.

Although not shown for brevity, we confirm that there is a good agreement between the sampled line-of-sight velocity dispersion profiles and the mock observational data for model C. Fig.~\ref{fig:MockModel3Cont} shows the marginalised posterior probability distribution of our fitting parameters for model~C with the cyan line with the cyan solid square representing the true values of the parameters. Except for $\log(m_{\rm DM})$ (that is now expected to be challenging to detect), the true parameter values are well recovered. 
 
Contrary to our MCMC results for the observational data (Fig.~\ref{fig:MODEL1CON}), the probability distribution of $\log(m_{\rm DM})$ does not show a clear degeneracy with $\log(\rho_{\rm b,NSC})$. The close-up view of the marginalised probability distribution of $\log(m_{\rm DM})$ is shown in Fig.~\ref{fig:MockModel3Hist}. Similar to model~A, Fig.~\ref{fig:MockModel3Hist} shows a clear gap between about $\log[m_{\rm DM}{\rm (eV)}]=-20.4$ and $-18.5$, unlike model~B that has a soliton core with $m_{\rm DM}=10^{-19.5}$~eV. Hence, we can confidently conclude that the gap can be used to reject a soliton core with ULDM particle mass in the range between $m_{\rm DM}=10^{-20.4}$~eV and $10^{-18.5}$~eV.
On the other hand, comparing with  model~A (Fig.~\ref{fig:MockModel1Hist}),  there is no clear peak of the probability distribution around $\log[m_{\rm DM}{\rm (eV)}]=-20.5$ in model~C. This means that the $10^{-20.5}$~eV ULDM particle mass is equally possible to be $m_{\rm DM}<10^{-21.0}$~eV or $m_{\rm DM}>10^{-18.5}$~eV. In other words,
the current quality of the data cannot identify or reject the ULDM particle mass outside of the gap, i.e. $m_{\rm DM}<10^{-20.4}$~eV or $m_{\rm DM}>10^{-18.5}$~eV, including $10^{-20.5}$~eV.

Interestingly, the fact that the result for the observational data (Fig.~\ref{fig:MODEL1Hist}) has a clear peak around $\log[m_{\rm DM}{\rm (eV)}]=-20.5$ indicates two potential scenarios: there is a soliton core with $m_{\rm DM}=10^{-20.5}$~eV, or there is an extra mass contribution, compared to the pure NSC model of model~C, to mimic the $m_{\rm DM}=10^{-20.5}$~eV soliton core. Since the former scenario requires an unreasonably small mass of NSC, as discussed above, we think that the latter scenario is likely, because the mass of the nuclear stellar disk might become significant around $\sim$ 3 pc \cite[]{gallego2018}.

\begin{figure*}
    \centering
    \includegraphics[width=\textwidth]{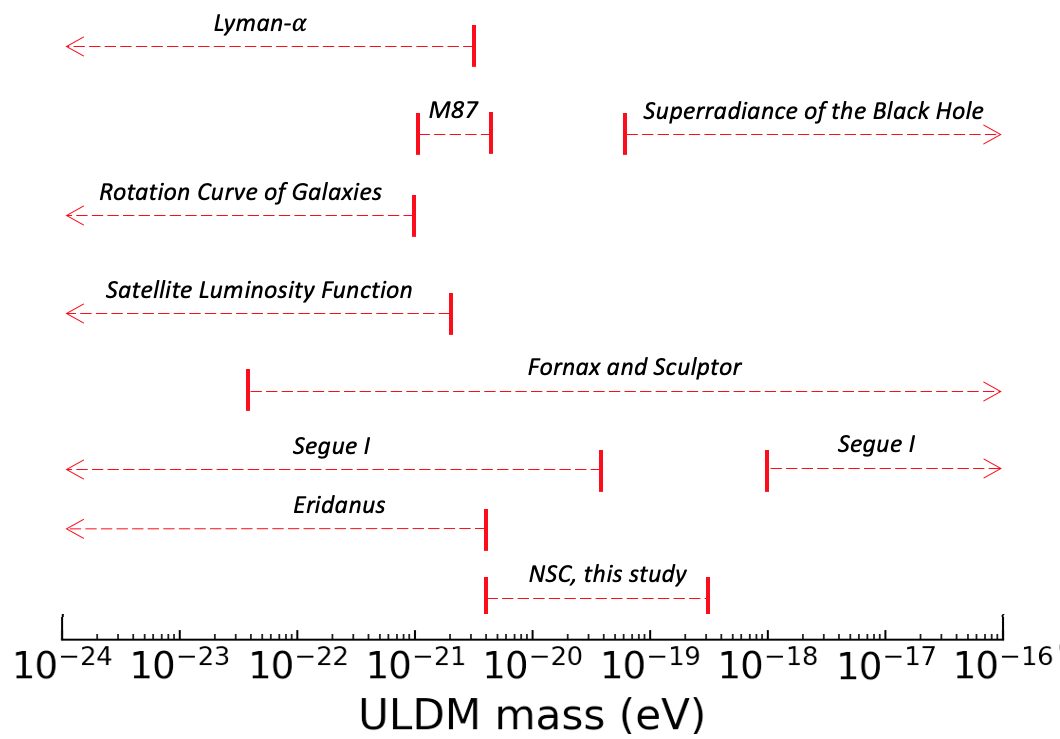}
    \caption{Summary of rejected ULDM particle masses from various astronomical probes. 
    The Lyman-$\alpha$ forest observation rejects $m_{\rm DM}<10^{-20.5}$~eV \citep{2017Ir, 2017Kobayashi, 2017Armengaud}. 
    The observed spin of black holes constrain the superradiance of black holes, and rejects $m_{\rm DM}>10^{-19.2}$~eV \citep{Stott+Marsh18}, including the Event Horizon Telescope observation of M87, which rejects $10^{-21.07}<m_{\rm DM}<10^{-20.34}$~eV \citep{2019Davoudiasl}. 
    Rotation curves of nearby galaxies also reject $m_{\rm DM}<10^{-21.0}$~eV \citep{2019Bar}. 
    \citet{2020Schutz} suggests that $m_{\rm DM}<10^{-20.7}$~eV is rejected by the satellite luminosity function inferred from the perturbed stellar streams \citep{2019Banik} and lensed images \citep{2020Gilman}, similarly to constraints on the WDM mass (Sec.~\ref{sec:intro}).
    \citet{Gonzalez-Morales+Marsh+Penarrubia+17} reject $m_{\rm DM}>10^{-22.4}$~eV from the stellar kinematics of the Fornax and Sculptor dwarf spheroidal galaxies. 
    \citet{2021Hayashi} find that the stellar kinematics of Segue~I is consistent with $10^{-19.4}<m_{\rm DM}<10^{-18.0}$~eV. 
    We naively take this as the required ULDM mass range, and consider that the other mass ranges are rejected, if the Segue~I stellar kinematics is purely due to the soliton core. \citet{2021Zoutendijk} reject $m_{\rm DM}<10^{-20.4}$~eV from the stellar kinematics of the ultra-faint dwarf galaxy, Eridanus. 
    }
    \label{fig:ULDMconstraints}
\end{figure*}

\section{Conclusions}
\label{sec:conclusion}
We have tested the existence of a soliton core due to Ultra-Light Dark Matter (ULDM) in the centre of the Milky Way by fitting the line-of-sight velocity dispersion data of its Nuclear Star Cluster (NSC) stars, taken from \citet{fritz2016}. We assumed a spherical isotropic Jeans model, using strong priors on the accurately measured NSC stellar number density profile and the mass of the SMBH. We fit the NSC density, $\rho_{\rm b,NSC}$, ULDM particle mass, $m_{\rm DM}$, the inner slope of the NSC density profile, $\gamma$, and the SMBH mass, $m_{\rm BH}$. The resultant marginalised probability distribution function of $m_{\rm DM}$ shows a peak around about $10^{-20.5}$~eV and a gap between about $10^{-20.4}$~eV and $10^{-18.5}$~eV, rejecting ULDM over this mass range. We show that this result is insensitive to our model assumptions and priors (see Appendices~\ref{sec:appendix} and \ref{sec:appendixB}). 
We also construct mock velocity dispersion data with the same radial bins and uncertainties as the observational data with different $m_{\rm DM}$, further validating our observational constraints. 

Fig.~\ref{fig:ULDMconstraints} shows a summary of the rejected ULDM mass ranges from a range of astronomical probes in the literature \citep[a comprehensive review can be found in][]{2021Hu}, including our new result. Taken at face value, Fig.~\ref{fig:ULDMconstraints} suggests that ULDM is not a viable solution for resolving the small scale problems in $\Lambda$CDM. Fig.~\ref{fig:ULDMconstraints} also highlights that our study provides a unique constraint on ULDM over a mass range only otherwise probed by the stellar kinematics of Milky Way satellite galaxies \citep[e.g.][]{Gonzalez-Morales+Marsh+Penarrubia+17,2021Hayashi}. 

However, there are four important caveats to our constraint. Firstly, We applied a spherical isotropic model for NSC. Applying an axisymmetric kinematic model, \citet{chatzopoulos2015} found a flatter NSC with $q=0.73\pm0.04$ and also suggested that a spherical model underestimates the total mass derived from the observed velocity dispersion profile. However, it requires a further study to address if a more realistic and complex model increases the NSC mass or provides more room for the ULDM soliton core. Secondly, we assumed that there is no radial dependence of the mass-to-light ratio. To some degree, the inner density slope parameter of $\gamma$ captures such radial dependence. However, this also requires further investigation in a future study. Thirdly, we have assumed throughout a single ULDM partilce comprises all of the dark matter. Finally, as highlighted in \citet{2020Davies}, a soliton core with $m_{\rm DM}>10^{-19.4}$~eV cannot survive in the Milky Way due to accretion into the SMBH. Hence, the stellar kinematics of the centre of the Milky Way may not be able to constrain the existence of a ULDM soliton core with $m_{\rm DM}>10^{-19.4}$~eV.


Constraining a ULDM mass lower than $10^{-20.0}$~eV with the methodology we introduce here would be still interesting, but require the stellar kinematic data at larger radii, $r>3$~pc. Further spectroscopic surveys of the stars in the Galactic centre with {\it VLT/KMOS} \citep[e.g.][]{Fritz+Patrick+Feldmeier-Krause+20} and future {\it VLT/MOONS} and {\it Subaru/ULTIMATE} would be invaluable to test the existence of the ULDM with $m_{\rm DM}<10^{-20.0}$~eV. In addition, the {\it Japan Astrometry Satellite Mission for INfrared Exploration} \citep[{{\it JASMINE};}][]{Gouda2012,Gouda+2020}\footnote{
\url{http://jasmine.nao.ac.jp/index-en.html}} will provide near-infrared astrometry for bright stars in the Galactic centre, which would provide further constraints on ULDM. This will require accurately modelling the nuclear stellar disc dynamics, since at $r>3$~pc the nuclear stellar disc dominates the central potential over the NSC \citep[e.g.][]{Li+Shen+Schive20}.

\section*{Data Availability}

The data underlying this article will be shared on reasonable request to the corresponding author. 

\section*{Acknowledgements}
FT, DK and GS acknowledge the support of the UK's Science \& Technology Facilities Council (STFC grant ST/N000811/1 and doctoral training grant ST/T506485/1). This research has made use of the VizieR catalogue access tool, CDS, Strasbourg, France \citep{Ochsenbein+00}.




\bibliographystyle{mnras}
\bibliography{bibliography} 




\appendix

\section{Systematic uncertainty of the black hole mass}
\label{sec:appendix}

\begin{figure}
    \centering
    \includegraphics[width = \columnwidth]{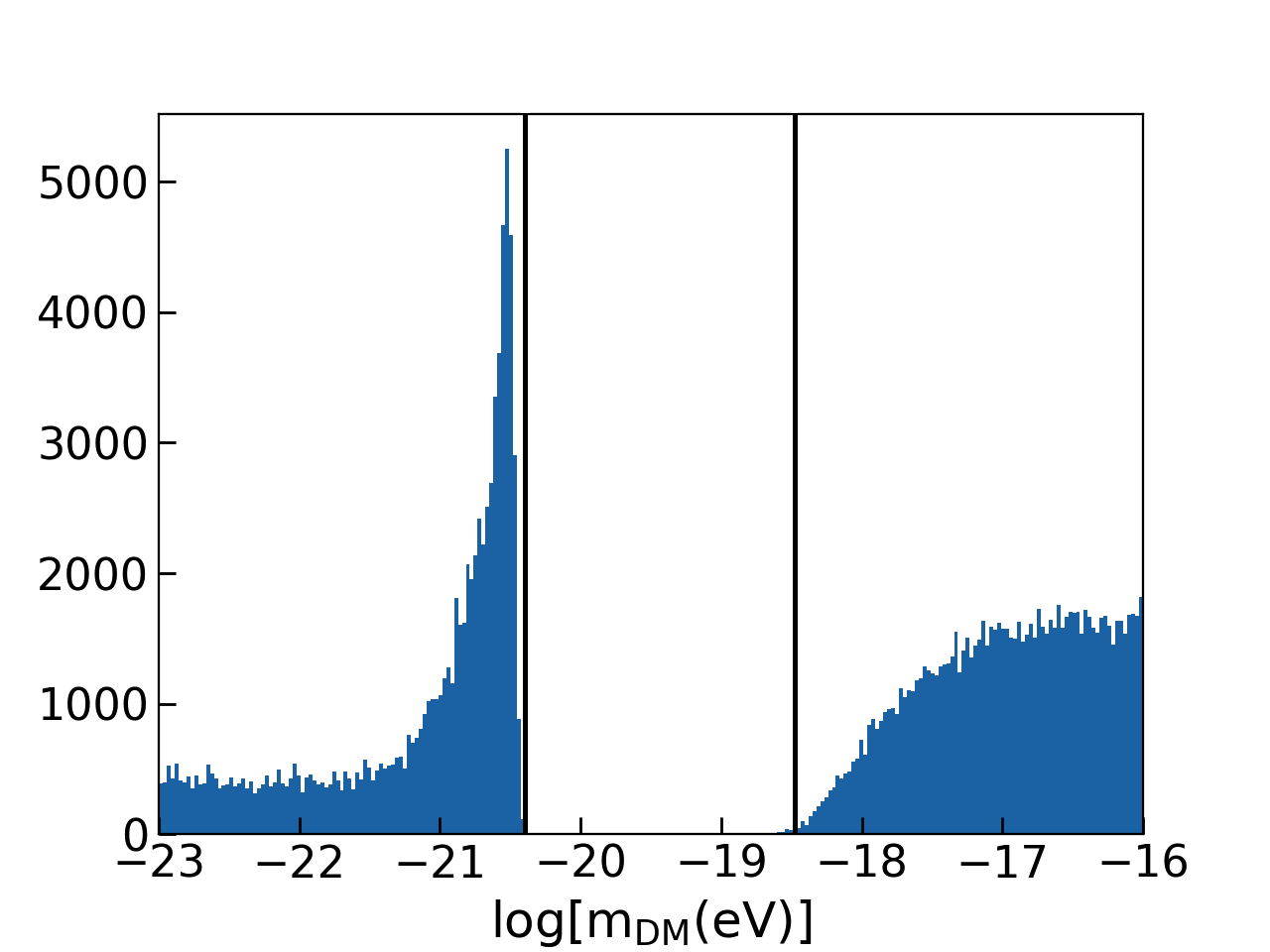}
    \caption{Marginalised posterior probability distribution of the model parameter $\rm log(m_{\rm DM})$ for lower black hole mass case. The marginalised posterior probability distribution is divided in to 250 bins. Solid black line indicates $\log[m_{\rm DM}{\rm (eV)}]=-20.4$ and $-18.5$.}
    \label{fig:DiffBHcase1Hist}
\end{figure}

\begin{figure}
    \centering
    \includegraphics[width = \columnwidth]{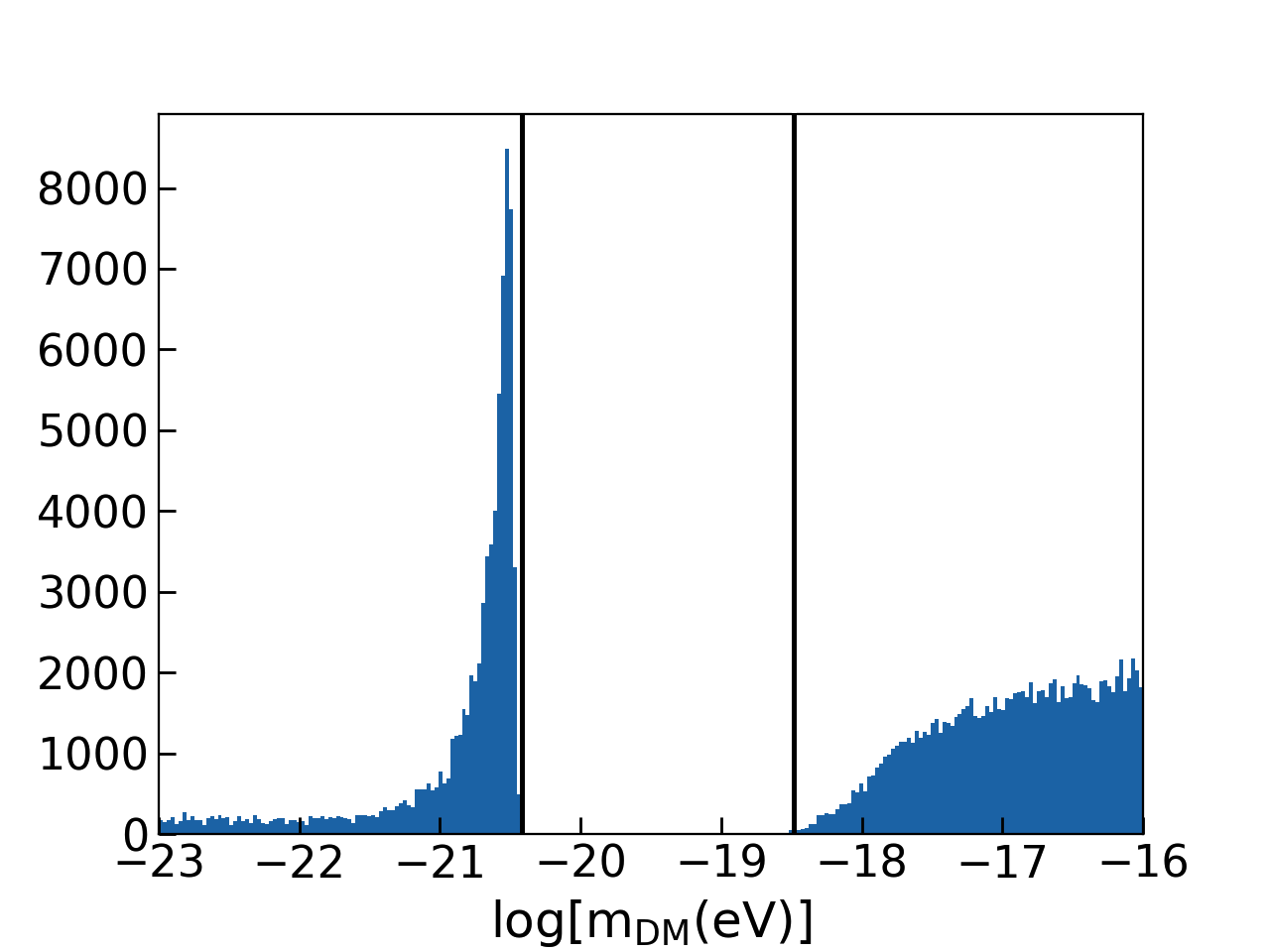}
    \caption{Marginalised posterior probability distribution of the model parameter $\rm log(m_{\rm DM})$ for higher black hole mass case. The marginalised posterior probability distribution is divided in to 250 bins. Solid black line indicates $\log[m_{\rm DM}{\rm (eV)}]=-20.4$ and $-18.5$.}
    \label{fig:DiffBHcase2Hist}
\end{figure}

There is a strong correlation between the distance to the Galactic centre, $R_{\rm 0}$, and $m_{\rm BH}$ measurements by \citet{abuter2020detection}, as shown in their Fig.~E2. \citet{abuter2020detection} estimate that there is a systematic uncertainty of 45\,pc for $R_{\rm 0}$, which propagates to a larger systematic uncertainty on the SMBH mass than the uncertainty considered in this paper. We tested the effect of this relatively large systematic uncertainty by considering two cases. The first case takes a distance to the Galactic centre of $R_{\rm 0}$ = 8.20 kpc, which is systematically shorter than our fiducial assumed distance. By fitting the correlation between $R_0$ and $m_{\rm BH}$ by eye from Fig.~E2 of \citet{abuter2020detection}, this corresponds to a SMBH mass of $m_{\rm BH}$ = 4.20 $\times 10^{6}$~$\rm M_{\odot}$. The different $R_0$ also affects the conversion of arcsec to pc, and we adjust the project radial distance of the stars from Sgr~A$^{*}$ and the break radius of the NSC density profile. The second case applies a larger distance to the Galactic centre of $R_{\rm 0}$ = 8.29 kpc. This leads to $m_{\rm BH} = 4.32 \times 10^{6}$~M$_{\odot}$. Figs.~\ref{fig:DiffBHcase1Hist} and \ref{fig:DiffBHcase2Hist} show the marginalised probability distribution of $\log(m_{\rm DM})$ for the former and latter cases, respectively, after fitting the data with the same method as in Section~\ref{sec:method}. These results show almost identical results to Fig.~\ref{fig:MODEL1Hist}. This confirms that the systematic uncertainty on $R_0$ and $m_{\rm BH}$ in \citet{abuter2020detection} is still small enough that it does not affect our conclusions.

\begin{figure}
    \centering
    \includegraphics[width = \columnwidth]{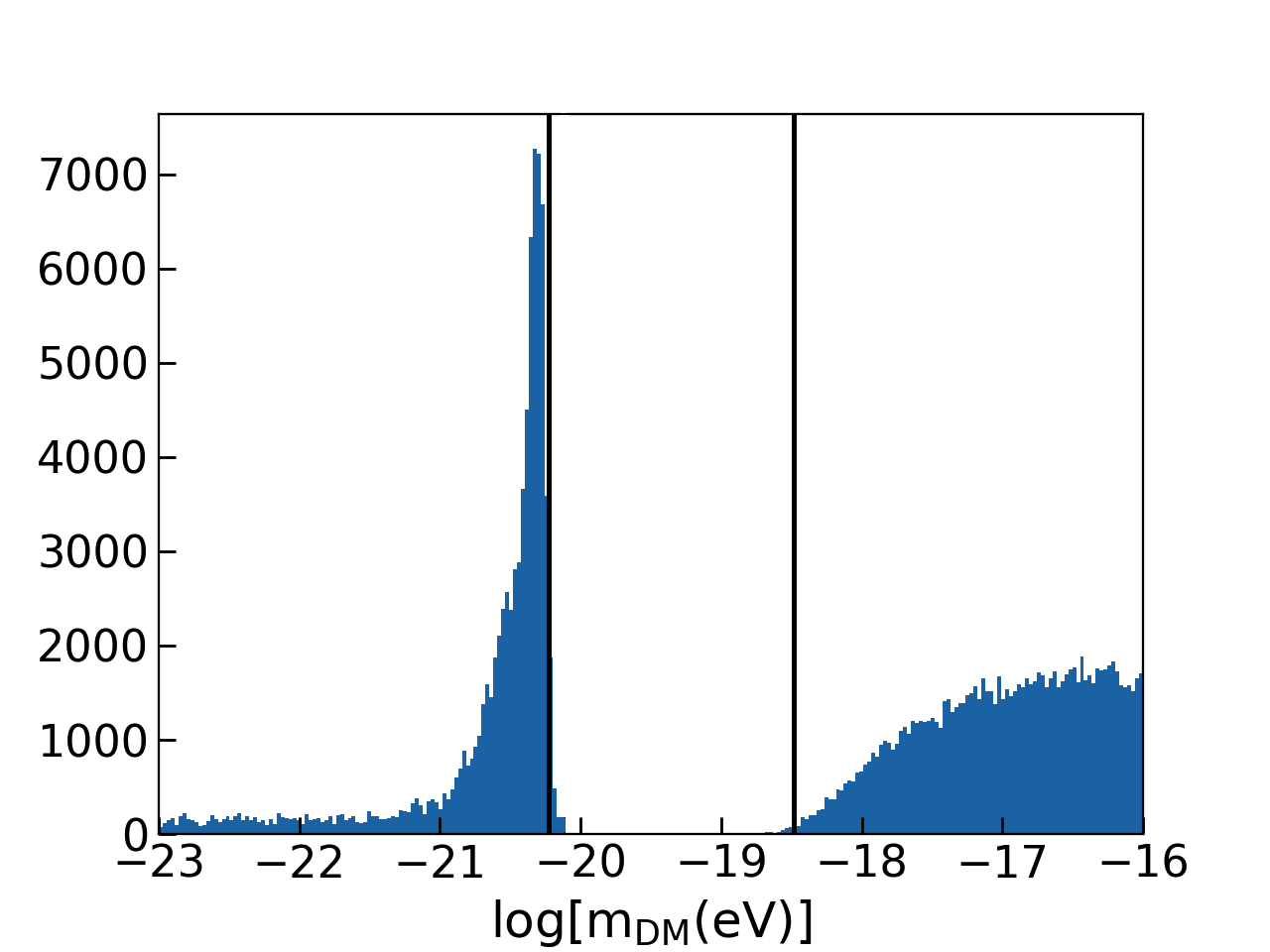}
    \caption{Marginalised posterior probability distribution of the model parameter $\rm log(m_{\rm DM})$ for the MCMC fitting result with a lower Milky Way mass of $M_{\rm h} = 9\times10^{11}$~$\rm M_{\sun}$, taken from \citet{2021Vasiliev}. 
    Solid black line indicates $\log[m_{\rm DM}{\rm (eV)}]=-20.4$ and $-18.5$.}
    \label{fig:MhVes}
\end{figure}

\section{The lower Milky Way mass case}
\label{sec:appendixB}

\citet{2021Vasiliev} recently suggest that the Milky Way's virial mass is as small as $9\times10^{11}$~M$_{\sun}$.
Fig.~\ref{fig:MhVes} shows the marginalised probability distribution of $\log(m_{\rm DM})$ obtained by the MCMC fitting to the observed velocity dispersion with adapting $M_{\rm h}=9\times10^{11}$~M$_{\sun}$. The result is similar to our fiducial result of Fig.~\ref{fig:MODEL1Hist} with $M_h=1.4\times10^{12}$~M$_{\sun}$, which is rather high side of the current estimates of the Milky Way mass. This demonstrates that our result is not sensitive to the assumed $M_{\rm h}$ value within the current expected range of $M_{\rm h}$ of the Milky Way.

\bsp	
\label{lastpage}
\end{document}